\documentclass[twocolumn,english]{revtex4}
\usepackage[T1]{fontenc}
\usepackage{color}
\usepackage{amsmath}
\usepackage{graphicx,epsf}
\usepackage{esint}
\usepackage{subfig}
\usepackage{mathrsfs}
\usepackage{epstopdf}
\usepackage{babel}


\begin{document}

\title{On nonlinear saturation of toroidal Alfv{\'e}n eigenmode due to thermal plasma nonlinearities}

\author{Ningfei Chen$^{1,2}$, Thomas Hayward-Schneider$^{1}$, Fulvio Zonca$^{2,3}$ ,Zhiyong Qiu$^{2,4}$, Zhixin Lu$^{1}$, Xin Wang$^{1}$, Alessandro Biancalani$^{5}$, Alexey Mishchenko$^{6}$,  Alberto Bottino$^{1}$, and Philipp Lauber$^{1}$\footnote{Author to whom correspondence should be addressed: philipp.lauber@ipp.mpg.de}}

\affiliation{$^1$ Max Planck Institute for Plasma Physics, Garching D-85748, Germany\\
$^2$Center for Nonlinear Plasma Science and   C.R. ENEA Frascati, C.P. 65, 00044 Frascati, Italy\\
$^3$Institute for  Fusion Theory and Simulation, School of Physics, Zhejiang University, Hangzhou 310027, China\\
$^4$Key Laboratory of Frontier Physics in Controlled Nuclear Fusion and Institute of Plasma Physics, Chinese Academy of Sciences,
Hefei 230031, China\\
$^5$De Vinci Higher Education, De Vinci Research Center, 92916 Paris, France\\
$^6$Max Planck Institute for Plasma Physics, Greifswald D-17491, Germany
}

\begin{abstract}
  The nonlinear saturation of toroidal Alfv{\'e}n eigenmode (TAE) due to thermal plasma nonlinearities is investigated using gyrokinetic particle-in-cell simulations and theoretical analysis. In the single toroidal mode number simulations with zonal fields filtered out, we find that the saturation level of TAE is governed by thermal plasma nonlinearities for $\gamma_{\rm L}/\omega_n>0.47\%$, which has weak dependence on the linear drive $\gamma_{\rm L}$, i.e., ``stiffness'' in saturation level. We find that the frequency of TAE decreases as the amplitude of it increases, which  is induced by the  phase-space zonal structure (PSZS) of thermal plasmas universally existed in particle-in-cell simulations. The saturation of TAE can be finally reached when the mode merges into the continuum. Following this process, the separation of neighboring poloidal harmonics and mode transition to energetic particle modes can be observed. In simulations with zonal fields, zonal fields can essentially counteract the effects of PSZS of thermal plasmas, leading to roughly a factor of 2 enhancement of the TAE saturation level compared to the single toroidal mode number simulation, implying the necessity of including zonal modes in evaluating the saturation level of TAE.
\end{abstract}

\maketitle

\section{Introduction}\label{sec:Introduction}

The confinement and slowing-down of energetic particles (EPs) generated
by fusion reaction and/or neutral beam injection (NBI) are essential
topics in achieving self-sustained burning plasma \cite{LChenRMP2016}. Alfv{\'e}n instabilities,
which can be driven unstable by EPs \cite{LChenPoP1994}, can induce significant EP transport \cite{AFasoliNF2007},
among which the toroidal Alfv{\'e}n eigenmode (TAE) is an important
component due to its minimal continuum damping in toroidicity-induced
gaps \cite{CZChengAP1985,GFuPoFB1989}. Thus, the nonlinear saturation mechanism of TAE is an essential
topic. 

Previous studies mainly focused on the nonlinear saturation of TAE
due to EP phase space nonlinear dynamics, i.e., wave-particle trapping
theory proposed by Berk and Breizman \cite{HBerkPoFB1990a,HBerkPoFB1990b}, which predicts a quadratic
dependence of saturation level on the linear growth rate. Recent modelling work has taken into account the effect of the beat-driven zonal mode on the saturation level of the EP-driven modes \cite{barberis2025perturbative}. For future
tokamak devices, the thermal velocity of EPs is similar to the Alfv{\'e}n
velocity $v_{{\rm A}}\equiv B_{0}/\sqrt{4\pi n_{{\rm i}}m_{{\rm i}}}$, leading to a  stronger resonance. Here, $B_0$ is the equilibrium magnetic field, $n_{\rm i}$ and $m_{\rm i }$ are the number density and mass of the thermal ion, respectively. Thus, a much larger linear drive and, consequently, a much higher saturation level are expected along this line. Another route towards nonlinear saturation of TAE is the wave-wave nonlinearity \cite{FZoncaPRL1995,LChenPPCF1998,ZQiuRMPP2023}. Among them, the beat-driven of $n=m=0$ zonal fields and phase-space zonal structure (PSZS) are of special interest \cite{FZoncaNJP2015}, because their generation bears no threshold on TAE amplitude and imposes various effects on TAE \cite{LChenPRL2012,LChenNF2025,QFangNF2025}. Here, $n/m$ are the toroidal/poloidal mode numbers. Zonal fields are zero-frequency toroidally symmetric perturbations, including electrostatic zonal flow and electromagnetic zonal current, which are accompanied by 
structures in the phase space of particles, i.e.,  PSZS,  generated by finite level fluctuations.  However, the PSZS  theory was originally designed to describe the nonlinear dynamics of EPs, since the redistribution of EPs  exhibits  a significant change in phase space. For thermal plasmas, their evolution in phase space was investigated with pressure gradient-driven drift wave (DW) dynamics \cite{LChenNF2007a,SWangPRL2024}, while their phase space evolution was hardly investigated in TAE nonlinear dynamics. Meanwhile, thermal plasma nonlinearities were found to affect the saturation level of TAE via ion-induced scattering \cite{LChenPoP2022,ZChengNF2024,ZChengNF2025}, but this mechanism requires a multi-$n$ simulation to evolve the spectrum of TAE. Recently, Ref. \cite{RMaarXiv2025} investigated the effect of thermal plasma nonlinearities on the nonlinear saturation of reversed shear Alfv{\'e}n eigenmode (RSAE), and a significant contribution can be found. 

In this work, we investigate the effects of thermal plasma nonlinearities on the nonlinear saturation of single-$n$ TAE using gyrokinetic theory and  particle-in-cell (PIC) simulations in the modified International Tokamak Physics Activity (ITPA) set of parameters.  More specifically, we show that thermal plasma nonlinearities contribute to the excitation of zonal field fluctuations as well as PSZS of thermal species, via beating of single-$n=6$ TAE and itself. Here, the individual contribution of zonal fields and PSZS is considered both from theory and simulation by retaining or filtering out $n=0$ modes. Meanwhile, PSZS  exists in the PIC simulation once the corresponding particle species evolves nonlinearly. The main findings are summarized below.
In the case without $n=0$ zonal fields, we find 

1. The saturation level of  TAE shows two regimes. For the linear drive below $\gamma_{\rm L}/\omega_n\simeq 0.47\%$, the nonlinear saturation is dominated by EP nonlinearity. Otherwise, the saturation of TAE is dominated by thermal plasma nonlinearity, i.e., PSZS of thermal species, which is $e\delta\phi_n/T_{\rm e}\sim0.1$ and  almost independent of linear drive, showing feature of ``stiffness''.
  Here, $\delta\phi_n$ and $\omega_n$ are the electrostatic potential and  real frequency of TAE with toroidal mode number $n=6$.

2. The saturation of TAE can happen when the mode amplitude reaches  $e\delta\phi_n/T_{\rm e}\sim0.1$ even if EPs evolve linearly, which implies a threshold on the amplitude of TAE imposed by thermal plasma nonlinearity.  Upon mode saturation, the decrease of mode frequency and the separation of $m=10$ and $m=11$ poloidal harmonics can be observed.

3. The decrease of mode frequency results from the modification of the potential well of TAE by PSZS of thermal species, as indicated by gyrokinetic theory developed herein. 
As the mode merges to the continuum, the mode saturation and separation of neighboring poloidal harmonics can be identified.
Meanwhile, the saturation level  can be quantitatively derived by the gyrokinetic theory via balancing the linear toroidal coupling and nonlinear decoupling induced by PSZS of thermal species. 

In the case with $n=0$ zonal field, we find

1. For reasonable linear drive $\gamma_{\rm L}/\omega_n>1\%$, the saturation of TAE is also dominated by thermal plasma nonlinearities. However, in this case, the effect of PSZS on TAE is significantly counteracted, leading to enhancement of the saturation level of TAE from $e\delta\phi_n/T_{\rm e}\sim0.1$ to $e\delta\phi_n/T_{\rm e}\sim0.2$.

2. For both cases with/without zonal fields, the saturation level of TAE is found to be proportionate to the square root of inverse aspect ratio, which is predicted by the theory. This implies a stronger TAE activity in devices with a larger inverse aspect ratio.

The results given above imply that for future tokamaks with much stronger EP drive, the saturation of TAE might be dominated by thermal plasma nonlinearity, leading to a much lower saturation level than that expected from EP nonlinearity. Meanwhile, it is necessary to keep $n=0$ zonal fields to evaluate the saturation level of TAE in PIC simulations, or it will be underestimated due to the universal existence of PSZS of thermal species.

The paper is organized as follows. In section \ref{sec:model}, the gyrokinetic model implemented in the code is introduced. In section \ref{sec:Linear-results}, the linear results of $n=6$ TAE are presented, with the dependence of mode frequency and growth rate on EP concentration and temperature being investigated. In sec. \ref{sec:Nonlinear-results}, the nonlinear saturation of TAE is investigated for both cases with and without $n=0$ zonal fields. In section \ref{sec:theory}, a gyrokinetic theory is developed to explain the simulation findings. Finally, summary and discussion are presented in section \ref{sec:discussion}.

\section{Simulation model}\label{sec:model}

The gyrokinetic Vlasov-Maxwell equations are solved in the $\delta f$
gyrokinetic PIC code ORB5 \cite{ELantiCPC2020}, with the distribution function
of species $s$ being separated into equilibrium/background component
$F_{0{\rm s}}$ and time-dependent perturbed distribution function
$\delta f_{{\rm s}}$, i.e., $f_{{\rm s}}\equiv F_{0{\rm s}}+\delta f_{{\rm s}}$.
Here, the subscripts $s={\rm i,e,EP}$ represent thermal ions, electrons and
energetic particles (EPs). The perturbed distribution function $\delta f_{{\rm s}}$
can be solved using the gyrokinetic equation
\begin{eqnarray}
 & \dfrac{\partial\delta f_{{\rm s}}}{\partial t}+\dot{\boldsymbol{R}}\cdot\left.\dfrac{\partial\delta f_{{\rm s}}}{\partial\boldsymbol{R}}\right|_{v_{\parallel}}+\dot{v}_{\parallel}\dfrac{\partial\delta f_{{\rm s}}}{\partial v_{\parallel}}\nonumber \\
 & =\delta\dot{\boldsymbol{R}}\cdot\left.\dfrac{\partial F_{0{\rm s}}}{\partial\boldsymbol{R}}\right|_{\varepsilon}-\delta\dot{\varepsilon}\dfrac{\partial F_{0{\rm s}}}{\partial\varepsilon}.\label{eq:gk_vlasov}
\end{eqnarray}

Here, $\dot{\boldsymbol{R}}=\dot{\boldsymbol{R}}_{0}+\delta\dot{\boldsymbol{R}}$,
$\dot{v}_{\parallel}=\dot{v}_{\parallel0}+\delta\dot{v}_{\parallel}$
represents gyrocenter trajectory, with $\left[\dot{\boldsymbol{R}}_{0},\dot{v}_{\parallel0}\right]$
being the unperturbed trajectory and $\left[\delta\dot{\boldsymbol{R}},\delta\dot{v}_{\parallel}\right]$
being the perturbed trajectory. The mixed-variable scheme is implemented
in ORB5 to solve the cancellation problem \cite{AMishchenkoPoP2014,AMishchenkoCPC2019}, in which the perturbed vector potential is separated into Hamiltonian and symplectic parts,
i.e., $\delta A_{\parallel}\equiv\delta A_{\parallel}^{{\rm h}}+\delta A_{\parallel}^{{\rm s}}$.
Then, the perturbed equations of motion are given by 
\begin{eqnarray*}
\delta\dot{\boldsymbol{R}} & = & \dfrac{\boldsymbol{b}}{B_{\parallel}^{*}}\times\nabla\left\langle \delta\phi-v_{\parallel}\delta A_{\parallel}^{{\rm s}}-v_{\parallel}\delta A_{\parallel}^{{\rm h}}\right\rangle \\
 &  & -\dfrac{q_{{\rm s}}}{m_{{\rm s}}}\left(\boldsymbol{b}_{0}^{*}+\dfrac{1}{B_{\parallel}^{*}}\left\langle \delta A_{\parallel}^{{\rm s}}\right\rangle \nabla\times\boldsymbol{b}\right)\left\langle \delta A_{\parallel}^{{\rm h}}\right\rangle ,
\end{eqnarray*}
\begin{eqnarray*}
\delta\dot{v}_{\parallel} & = & -\left(\dfrac{v_{\parallel}}{B_{\parallel}^{*}}\nabla\times\boldsymbol{b}+\dfrac{q_{{\rm s}}}{m_{{\rm s}}}\dfrac{1}{B_{\parallel}^{*}}\nabla\left\langle \delta A_{\parallel}^{{\rm s}}\right\rangle \times\boldsymbol{b}\right)\\
 &  & \cdot\nabla\left\langle \delta\phi-v_{\parallel}\delta A_{\parallel}^{{\rm h}}\right\rangle +\dfrac{q_{{\rm s}}}{m_{{\rm s}}}v_{\parallel}\boldsymbol{b}\cdot\nabla\left\langle \delta A_{\parallel}^{{\rm h}}\right\rangle \\
 &  & -\mu\dfrac{\boldsymbol{b}\times\nabla B}{B_{\parallel}^{*}}\cdot\nabla\left\langle \delta A_{\parallel}^{{\rm s}}\right\rangle ,
\end{eqnarray*}
\begin{eqnarray*}
\delta\dot{\varepsilon} & = & v_{\parallel}\delta\dot{v}_{\parallel}+\mu\delta\dot{\boldsymbol{R}}\cdot\nabla B.
\end{eqnarray*}

Here, $\delta\phi$ and $\delta A_{\parallel}$ are the perturbed
scalar and vector potential, $\boldsymbol{b}^{*}=\boldsymbol{b}_{0}^{*}+\nabla\left\langle \delta A_{\parallel}^{{\rm s}}\right\rangle \times\boldsymbol{b}/B_{\parallel}^{*}$,
$\boldsymbol{b}_{0}^{*}\approx\boldsymbol{b}+m_{{\rm s}}v_{\parallel}\nabla\times\boldsymbol{b}/\left(q_{{\rm s}}B_{\parallel}^{*}\right)$,
$B_{\parallel}^{*}=\boldsymbol{b}\cdot\nabla\times\boldsymbol{A}^{*}$,
$\boldsymbol{A}^{*}=\boldsymbol{A}+\left(m_{{\rm s}}v_{\parallel}/q_{{\rm s}}\right)\boldsymbol{b}$
is the modified vector potential, $\boldsymbol{B}=\nabla\times\boldsymbol{A}$,
$\boldsymbol{b}\equiv\boldsymbol{B}/B$ is the unit vector in the
direction of equilibrium magnetic field, $q_{{\rm s}}$ and $m_{{\rm s}}$
are the charge and mass of particle species, $\mu=v_{\perp}^{2}/\left(2B\right)$
is the magnetic momemnt, $\varepsilon=v_{\parallel}^{2}/2+\mu B$
is the kinetic energy of particles. $\left\langle \cdots\right\rangle $
represents the gyro-average. In the drift-kinetic limit adopted in
this paper, the gyroaverage is absent. The symplectic part of the
perturbed vector potential is defined such that 
\begin{eqnarray*}
\dfrac{\partial}{\partial t}\delta A_{\parallel}^{{\rm s}}+\boldsymbol{b}\cdot\nabla\delta\phi & = & 0.
\end{eqnarray*}

Meanwhile, the Hamiltonian part can be obtained from the parallel
Ampere's law
\begin{eqnarray*}
\left(\sum_{{\rm s}}\dfrac{\beta_{{\rm s}}}{\rho_{{\rm s}}^{2}}-\nabla_{\perp}^{2}\right)\delta A_{\parallel}^{{\rm h}} & = & \mu_{0}\sum_{{\rm s}}\delta j_{\parallel{\rm s}}+\nabla_{\perp}^{2}\delta A_{\parallel}^{{\rm s}}.
\end{eqnarray*}

Here, $\delta j_{\parallel{\rm s}}$ is the perturbed parallel gyrocenter
current, $\rho_{{\rm s}}\equiv\sqrt{m_{{\rm s}}T_{{\rm s}}}/\left(q_{{\rm s}}B\right)$,
$\beta_{{\rm s}}\equiv\mu_{0}n_{{\rm s}}T_{{\rm s}}/B^{2}$ is the ratio
between the thermal pressure of species $s$ and the magnetic pressure. The
electrostatic potential can be obtained from the quasi-neutrality
condition
\begin{eqnarray*}
-\nabla\cdot\sum_{{\rm s=i,EP}}\dfrac{q_{{\rm s}}^{2}n_{{\rm s}}}{T_{{\rm s}}}\nabla_{\perp}\delta\phi & = & \sum_{{\rm s}}q_{{\rm s}}\delta n_{{\rm s}}.
\end{eqnarray*}

For each particle species, if $\delta\dot{\boldsymbol{R}}=0$ and
$\delta\dot{v}_{\parallel}=0$ are imposed to the left hand side of
equation (\ref{eq:gk_vlasov}), the corresponding particle species
will follow the unperturbed trajectory $\left[\dot{\boldsymbol{R}}_{0},\dot{v}_{\parallel0}\right]$,
and the linear gyrokinetic equation is solved. In the following context,
we will refer to this particle species with ``linear'' treatment. Otherwise, the particles
follow the full orbit and the nonlinear gyrokinetic equation
with both parallel and perpendicular nonlinearities is solved, which
will be referred to as ``nonlinear''.

The perpendicular nonlinearity is given by the term $\delta\dot{\boldsymbol{R}}\cdot\partial\delta f_{{\rm s}}/\partial\boldsymbol{R}$.
The parallel nonlinearity is given by the term $\delta\dot{v_{\parallel}}\partial\delta f_{{\rm s}}/\partial v_{\parallel}$
in equation (\ref{eq:gk_vlasov}). Thus, it can be turned off by imposing
$\delta\dot{v}_{\parallel}=0$ in the code.

\section{Simulation setup}\label{sec:Simulation-setup}

The International Tokamak Physics Activity (ITPA) TAE case is employed
in this work, in which a large aspect ratio tokamak configuration
with major and minor radius being $R=10{\rm m}$ and $a=1{\rm m}$
is implemented \cite{konies2018benchmark}. The equilibrium toroidal magnetic field is $B_{0}=3{\rm T}$.
The safety factor is $q\left(s\right)=1.71+0.16s^{2}$ in the standard ITPA
case, where $s\equiv\sqrt{\psi/\psi_{{\rm a}}}$ is the normalized radial
coordinate used in ORB5, $\psi$ is the poloidal magnetic flux divided by $2\pi$, and
$\psi_{{\rm a}}$ is $\psi$ at plasma edge. The corresponding shear
Alfv{\'e}n continuum is shown in Figure \ref{fig:Density_q_profile}(a).
The number density of thermal ion and electron at the reference position $s=s_{0}=0.5$ are $n_{0{\rm i}}=n_{0{\rm e}}=2\times10^{13}{\rm cm}^{-3}$,
and their temperatures are uniform in space, with $T_{{\rm i}}=T_{{\rm e}}=4{\rm keV}$. The electron
beta at axis is $\beta_{{\rm e}}\equiv8\pi n_{{\rm e}}T_{{\rm e}}/B_{0}^{2}=8.064\times10^{-3}$.
In this set of parameters, the upper and lower accumulation point
of $n=6$ TAE are $\omega_{{\rm U}}=0.300\omega_{{\rm A}0}=4.401\times10^{5}{\rm rad/s}$
and $\omega_{{\rm L}}=0.273\omega_{{\rm A}0}=4.005\times10^{5}{\rm rad/s}$,
where $\omega_{{\rm A}0}\equiv v_{{\rm A}}/R=1.467\times10^{6}{\rm rad/s}$
is the Alfv{\'e}n frequency used for normalization, and the Alfv{\'e}n velocity
is $v_{{\rm A}}=1.467\times10^{9}{\rm cm/s}$. To avoid confusion in
the following analysis, there is a finite thermal ion density gradient in this work,
while the thermal ion profile is flat in a typical ITPA case. 

The linear driving source for TAE is energetic
particle species with Maxwellian distribution, whose temperature
is denoted by $T_{{\rm EP}}$ and the density profile is given by 
\begin{eqnarray*}
\dfrac{n_{{\rm EP}}}{n_{{\rm EP}}\left(s_{0}\right)} & = & \exp\left(-\Delta\kappa_{n}\tanh\left(\dfrac{s-s_{0}}{\Delta}\right)\right),
\end{eqnarray*}
where $\Delta=0.2$ is the width of EP profile, and $\kappa_{n}$
is the density gradient of EPs at the reference point $s_{0}$. The electron density profile is determined such that the equilibrium charge quasi-neutrality
condtion is satisfied, i.e., $n_{{\rm e0}}=n_{{\rm i0}}+n_{{\rm EP}}$.
The corresponding plasma density gradient is shown in Figure \ref{fig:Density_q_profile}(b).

In the following, the simulations are performed in the drift-kinetic limit
to simplify the system to get a better correspondence between
simulation and theory, since there is only a beat-driven zonal field being
involved, while spontaneous excitation of zonal fields via modulational instability
may exist when gyro-averaging is considered. Note that both scalar and vector potentials are normalized to $e/T_{{\rm e}}$ in this paper. 

\begin{center}
\begin{figure}
\subfloat{\includegraphics[scale=0.26]{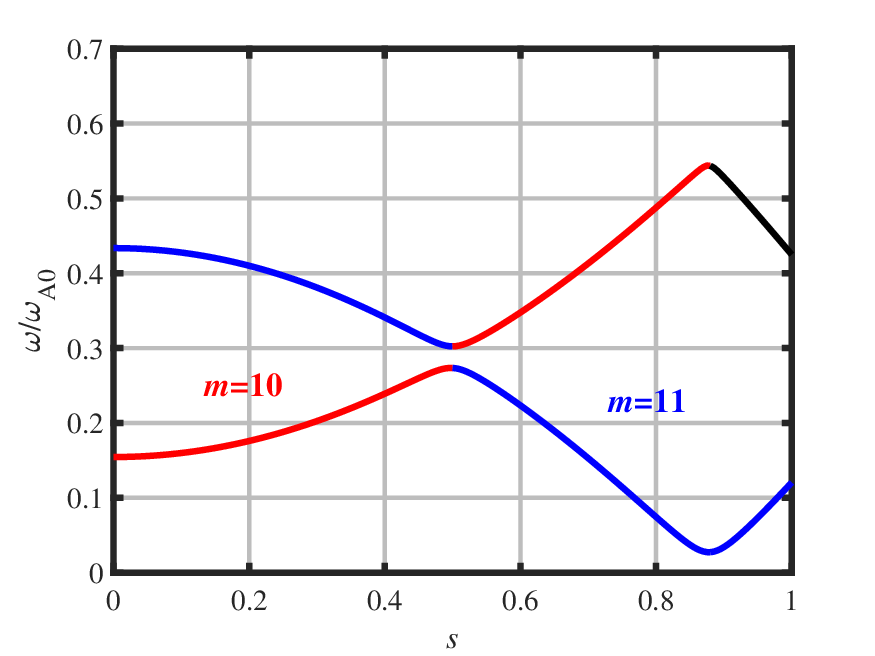}}\ \subfloat{\includegraphics[scale=0.26]{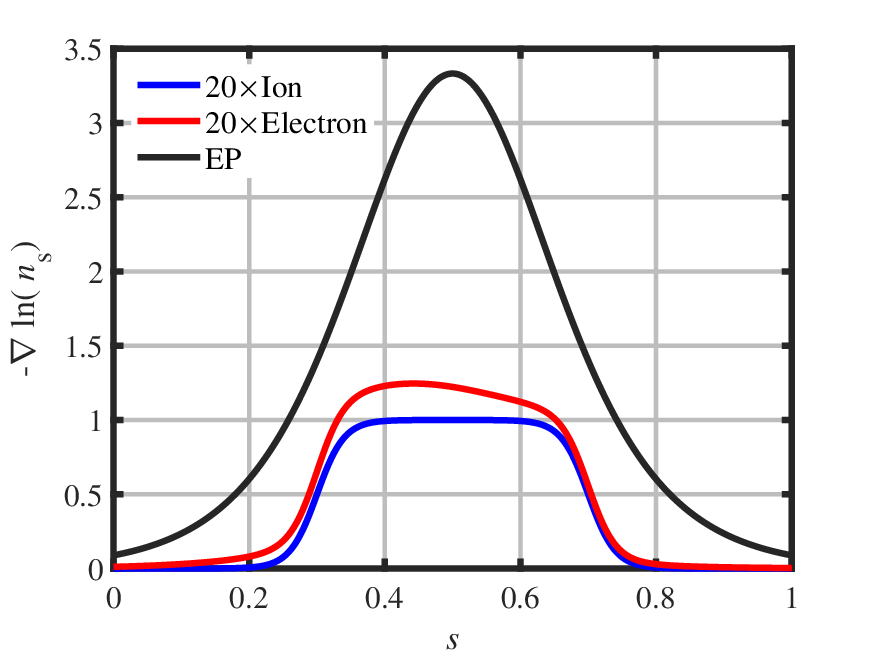}}\caption{(a) The Alfv{\'e}n continuum used in this paper, with $q\left(s\right)=1.71+0.16s^{2}$; (b) the profile of equilibrium logarithmic density gradient
of each species, with the black, red, and blue solid curves representing
that of EP, electron and thermal ion respectively. The gradient of
thermal ions and electrons is multiplied by 20 to get a better visualization.}\label{fig:Density_q_profile}
\end{figure}
\par\end{center}

\section{Linear results}\label{sec:Linear-results}

In this section, linear properties of TAE, e.g., growth rate and frequency,
are investigated, as the first step toward investigations on nonlinear
dynamics. The dependence of linear growth rate and frequency on EP
temperature is shown in Figure \ref{fig:growth_freq_EP_tmp}, with
the EP concentration being $n_{{\rm EP}}/n_{{\rm e}}=0.002$. It
is found that a maximum in linear growth rate is reached at $T_{{\rm EP}}\simeq500{\rm keV}$,
but it is not chosen as a default parameter in the following nonlinear
studies. Otherwise, $T_{{\rm EP}}=400{\rm keV}$ is the default EP
temperature, because the frequency at $T_{{\rm EP}}=500{\rm keV}$
is close to the upper continuum.

The dependence of the growth rate and the real frequency of TAE on EP concentration are shown in
Figure \ref{fig:growth_freq_EP_concentration}. It is found that both
the growth rate and frequency of TAE increase with increasing EP concentration.

The mode structures of poloidal harmonics of TAE are shown in Figure 
\ref{fig:radial_ms_phi_psi}. It is found that for both scalar and
vector potentials, the $m=10$ and $m=11$ poloidal harmonics
are dominant, which is a typical characteristic of TAE. Besides, we can also find that the ideal MHD constraint, i.e.,  $\delta\phi_{n}\approx \delta\psi_{n}$, is satisfied, which is also an important characteristic
of TAE. Here, $\delta\psi_{n}\equiv\omega_n\delta A_{||,n}/(k_{||}c)$ is the field quantity associated
with the magnetic perturbation of TAE, where $k_\parallel\equiv (nq-m)/(qR_0)$ is the parallel wavevector, $c$ is the light speed.
\begin{center}
\begin{figure}
\subfloat{\includegraphics[scale=0.25]{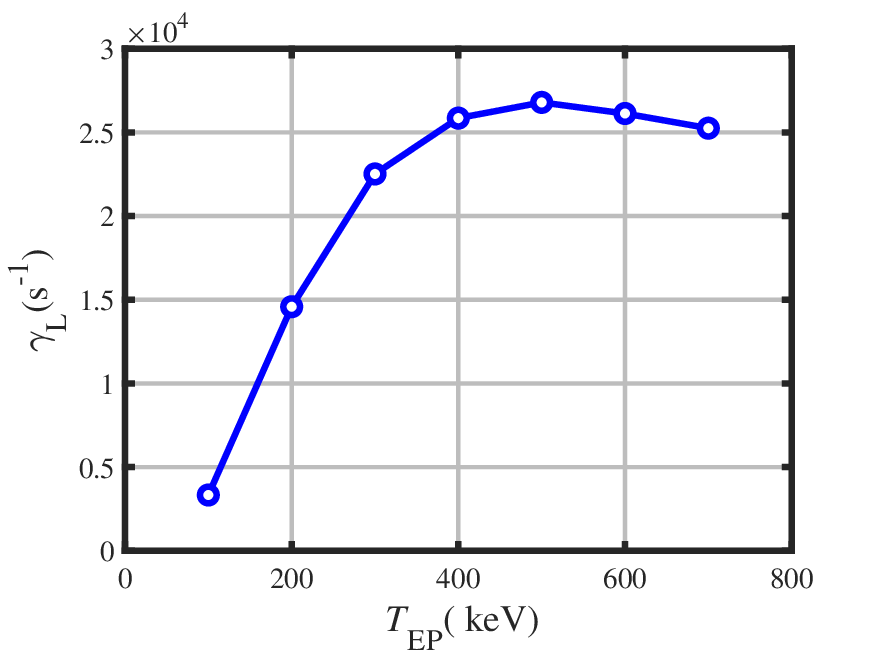}}\ \subfloat{\includegraphics[scale=0.25]{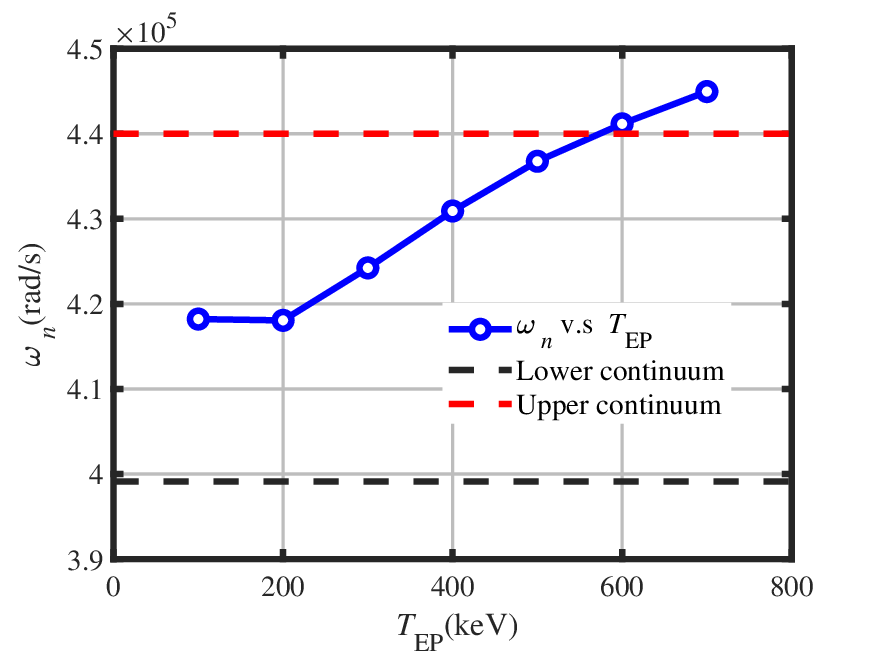}}

\caption{The dependence of the (a) growth rate, and (b) real frequency of TAE
on the EP temperature, with the EP concentration being $n_{{\rm EP}}/n_{{\rm e}}=0.002$.}\label{fig:growth_freq_EP_tmp}
\end{figure}
\par\end{center}

\begin{center}
\begin{figure}
\subfloat{\includegraphics[scale=0.25]{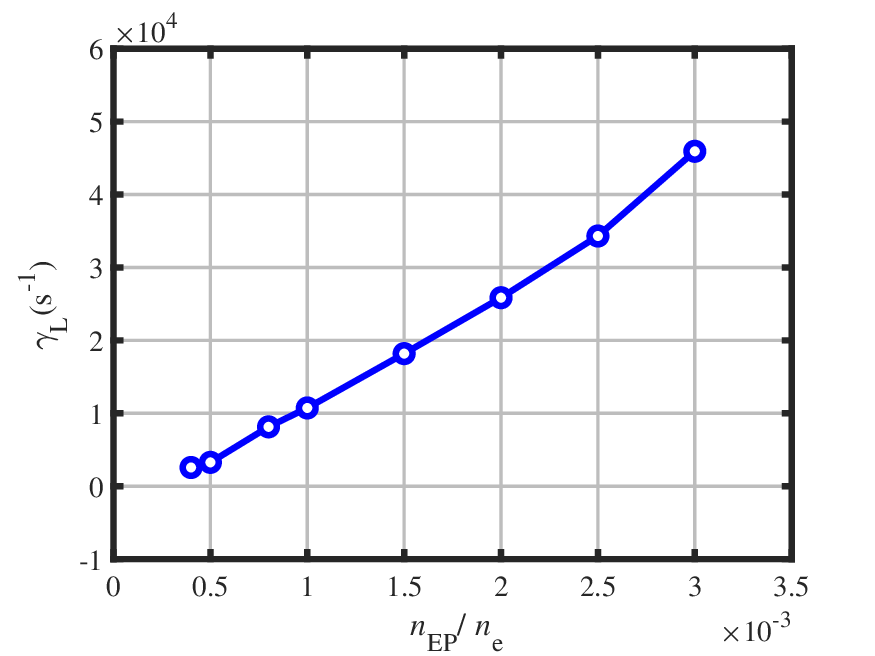}}\ \subfloat{\includegraphics[scale=0.25]{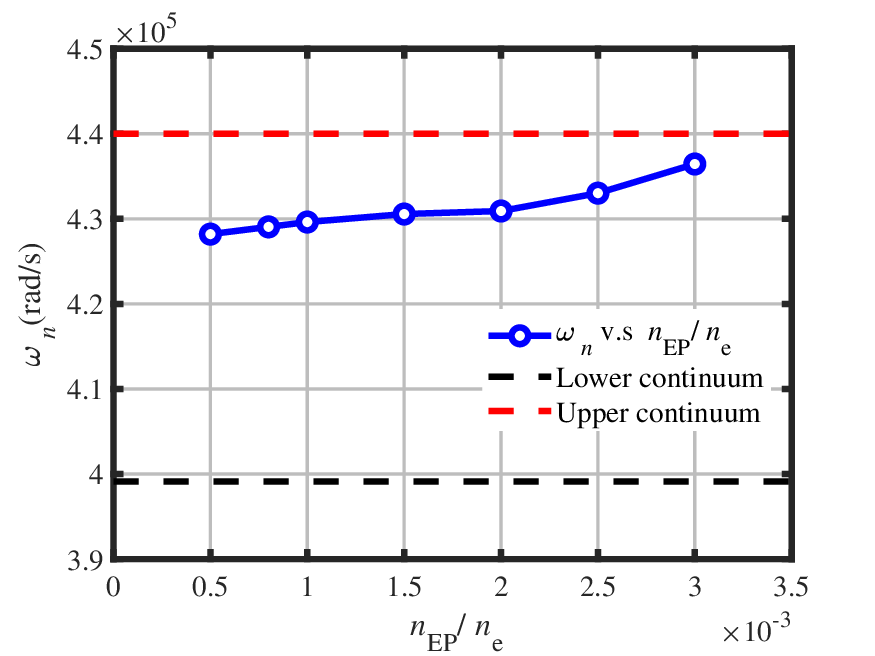}}

\caption{The dependence of the (a) growth rate, and (b) real frequency of TAE
on the EP concentration, with $T_{{\rm EP}}=400{\rm keV}$.}\label{fig:growth_freq_EP_concentration}
\end{figure}
\par\end{center}

\begin{center}
\begin{figure}
\subfloat{\includegraphics[scale=0.26]{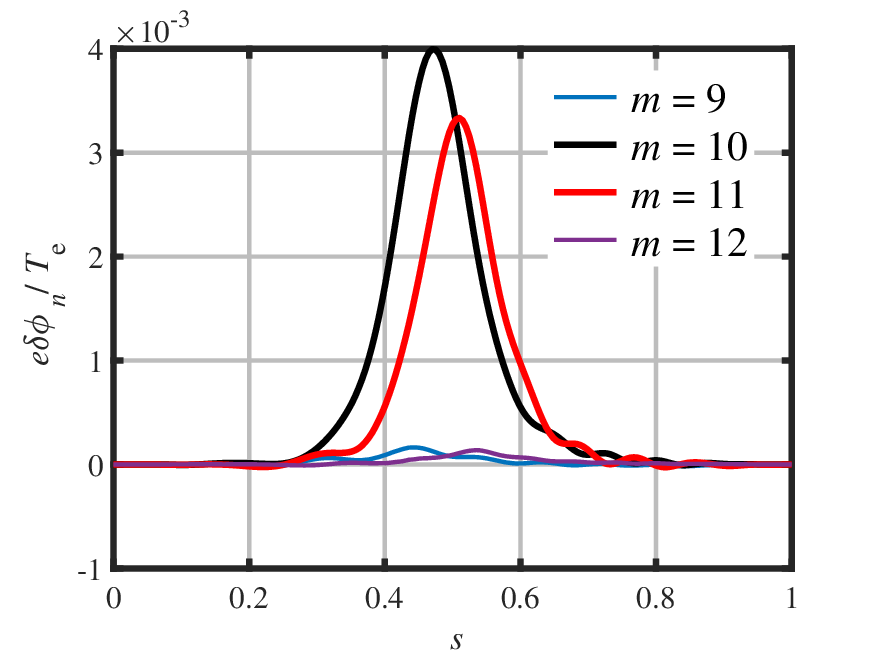}}\ \subfloat{\includegraphics[scale=0.26]{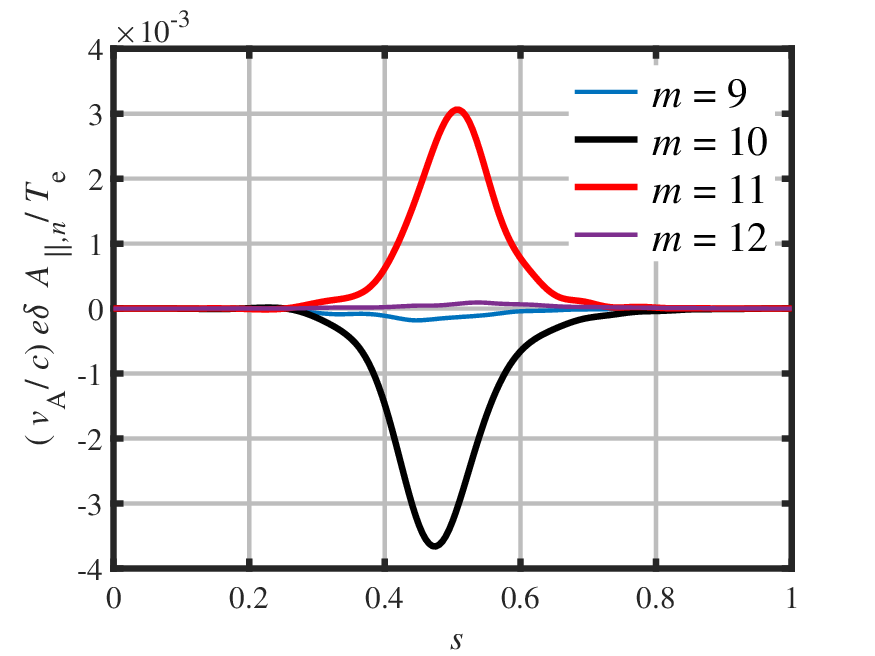}}\caption{The linear mode structure of four neighboring poloidal harmonics of (a) electrostatic
potential $e\delta\phi_n/T_{\rm e}$ and (b) normalized vector potential. Here,
$n_{{\rm EP}}/n_{{\rm e}}=0.002$, $T_{{\rm EP}}=400{\rm keV}$.}\label{fig:radial_ms_phi_psi}
\end{figure}
\par\end{center}

\section{Nonlinear results}\label{sec:Nonlinear-results}

In this section, we investigate the nonlinear dynamics of $n=6$ TAE
in both cases with and without $n=0$ modes. As mentioned in sec. \ref{sec:model}, there
is an option in the code that determines whether the corresponding particles of 
species `s' follow the perturbed or unperturbed trajectories, which will
be referred to as ``linear'' or ``nonlinear'' particle species in this context.
In previous studies, the wave-particle nonlinearity of EPs is a major
saturation mechanism for TAE, while the effects of thermal plasma nonlinearities on TAE dynamics and saturation will be analyzed in
this work. For simplicity, we would like to categorize our particle
settings into five cases, labeled with A, B, C, D, and E, as shown
in Table \ref{tab:cases}. In case A, all particle species are nonlinear.
Cases B, C, and D are used to emphasize the individual contribution
of electrons, thermal ions, and EPs, respectively. The case E indicates the impact of thermal plasma nonlinearity on the nonlinear dynamics of
TAE, with EPs evolving linearly. In the following, we will firstly reproduce the previous results
of the case with only EP nonlinearity, i.e., the case D. Then, the simulations with thermal plasma nonlinearities will be carried out
in both cases with and without $n=0$ zonal fields. According to the linear studies carried out above, the default value of EP density gradient $\kappa_{n}=3.333$, EP concentration $n_{\rm EP}/n_{\rm e}=0.002$, and EP temperature $T_{\rm EP}=400{\rm keV}$. Finally, according to the convergence study shown in the Appendix A, the mass ratio is $200$ for single-$n$ simulations, while the realistic mass ratio is used  for the simulation with $n=0$ modes.

\begin{center}
\begin{table}
\centering{}%
\begin{tabular}{|c|c|c|c|}
\hline 
Cases & thermal ion & electron  & EP\tabularnewline
\hline 
\hline 
A & NL & NL & NL\tabularnewline
\hline 
B & L & NL & L\tabularnewline
\hline 
C & NL & L & L\tabularnewline
\hline 
D & L & L & NL\tabularnewline
\hline 
E & NL & NL & L\tabularnewline
\hline 
\end{tabular}\caption{The definition of five cases used in this work. Here, L
and NL represent the corresponding species are ``linear'' and ``nonlinear'', respectively.}\label{tab:cases}
\end{table}
\par\end{center}

\subsection{Simulation results without zonal fields}

\subsubsection{Nonlinear saturation due to EP nonlinearity}\label{subsec:EP_nonlinearity}

Before investigating the nonlinear saturation of TAE with thermal
plasma nonlinearities, we need to identify the saturation due to EP nonlinearity \cite{ABiancalaniPPCF2017}, i.e., case D in Table \ref{tab:cases}. The temporal
evolution of $e\delta\phi_n/T_{\rm e}$ of $n=6$ TAE is shown in Figure 
\ref{fig:phimax_no_zf_0.002_EP_Br}. Here, we can observe a linear
growth before $\omega_{{\rm A0}}t\sim400$. After that, the growth
rate of TAE decreases with time, which finally yields a saturation
level $e\delta\phi_n/T_{\rm e}\sim 2$, or $\delta B_{r}/B_{0}\sim2\times10^{-3}$. The saturation
due to EP phase space nonlinearity is achieved when the bounce frequency
$\omega_{{\rm b}}$ balances the linear growth rate, i,e., $\omega_{{\rm b}}\sim\sqrt{\delta\phi_{n}}\sim3.2\gamma_{{\rm L}}$,
which finally yields a quadratic dependence of the saturation level
of TAE on the linear growth rate. The scaling law of saturation level on linear growth rate is given by red
squares in Figure \ref{fig:scaling_law_no_zf}, in which a quadratic
scaling law can be clearly observed. Note that the case shown in Figure 
\ref{fig:phimax_no_zf_0.002_EP_Br} corresponds to the rightmost red point
in Figure \ref{fig:scaling_law_no_zf}. Here, the linear drive is
altered by changing the EP concentration $n_{{\rm EP}}/n_{{\rm e}}$,
which aims to impose minimal influence on mode frequency. The $\phi(s,\omega)$
diagram of TAE in the nonlinear phase at $\omega_{{\rm A0}}t\sim700$
is shown in Figure \ref{fig:phi_s_omega_EP}. It is found that the
mode still exists in the TAE gap in the nonlinear saturation phase, with little separation of $m=10$ and $m=11$ harmonics observed.
This scaling provides a good foundation for the investigation with
thermal plasma nonlinearities, as we will investigate in the next.
\begin{center}
\begin{figure}
\begin{centering}
\includegraphics[scale=0.4]{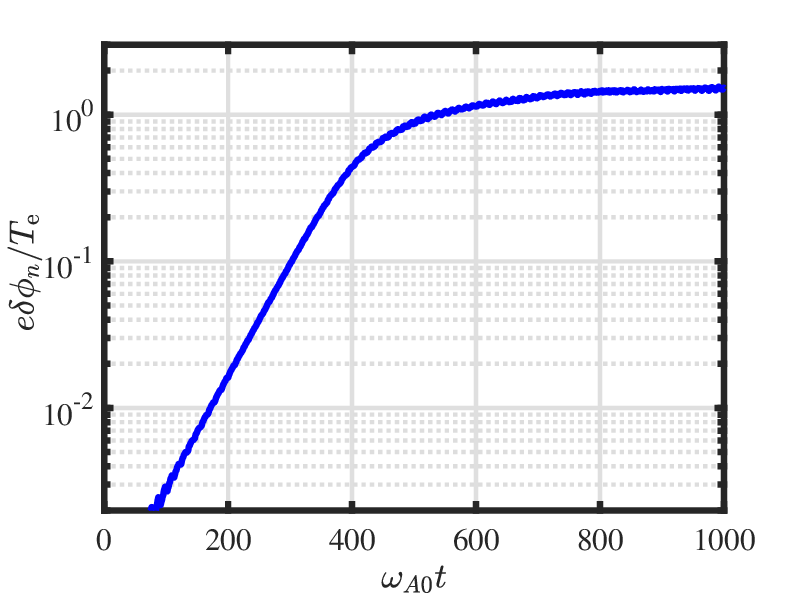}
\par\end{centering}
\caption{The temporal evolution of the maximum of the normalized electrostatic potential
$e\delta\phi_n/T_{\rm e}$ of TAE for $n_{{\rm EP}}/n_{{\rm e}}=0.002$
and $T_{{\rm EP}}=400{\rm keV}$. Note that this case corresponds to the rightmost red point in Figure \ref{fig:scaling_law_no_zf}.}\label{fig:phimax_no_zf_0.002_EP_Br}
\end{figure}
\par\end{center}

\begin{center}
\begin{figure}
\includegraphics[scale=0.5]{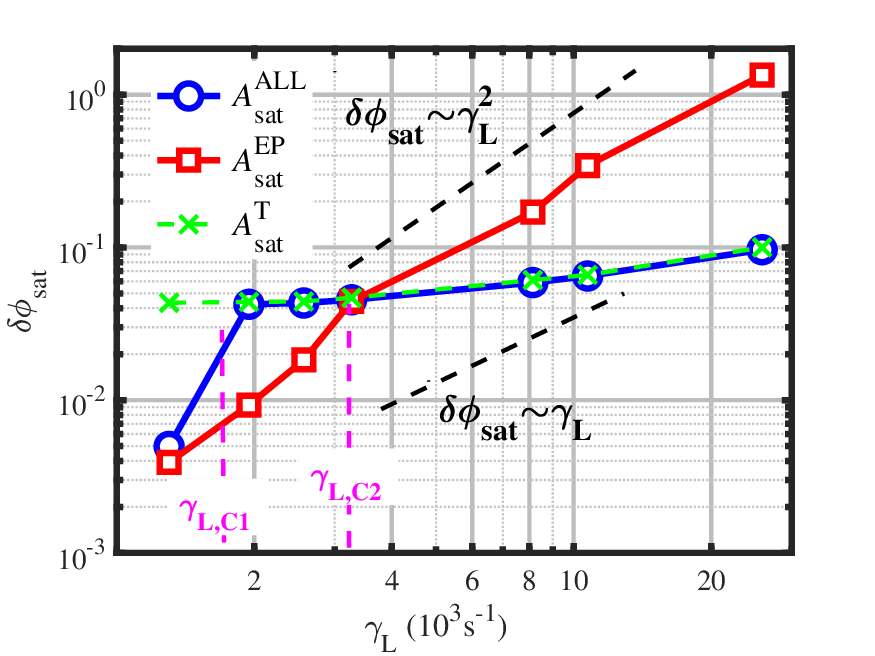}

\caption{The scaling law of the saturation of TAE versus the linear growth
rate $\gamma_{{\rm L}}$ in single-$n=6$ simulation. The blue circles,
red squares, and green crosses represent the scaling law for cases
A: with all nonlinearities $A_{\rm sat}^{\rm ALL}$, D: with only EP nonlinearity $A_{\rm sat}^{\rm EP}$, and E: with only thermal plasma nonlinearity $A_{\rm sat}^{\rm T}$. Two black lines are given as reference lines for quadratic                                          
and linear scaling.}\label{fig:scaling_law_no_zf}
\end{figure}
\par\end{center}

\begin{center}
\begin{figure}
\begin{centering}
\includegraphics[scale=0.38]{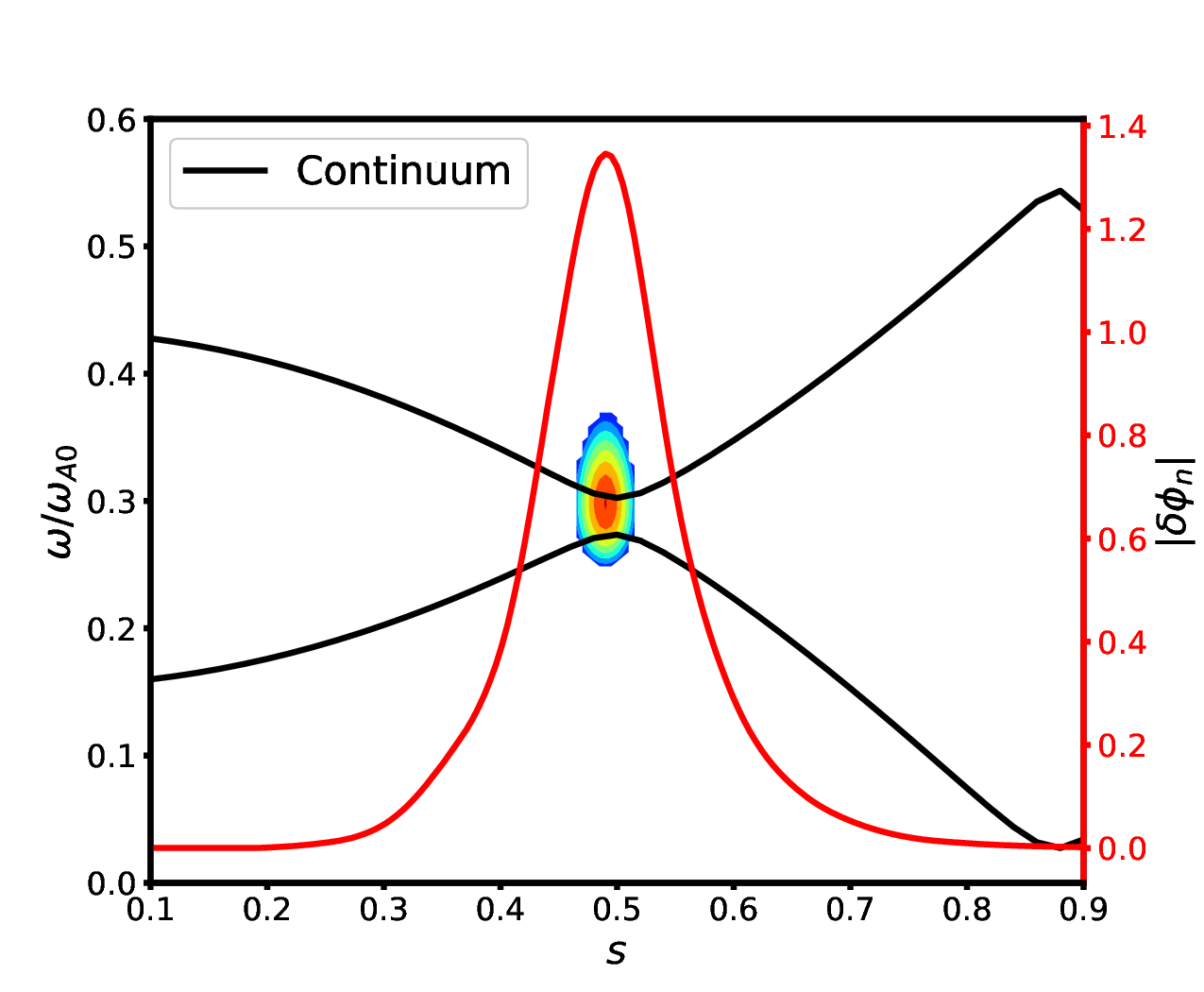}
\par\end{centering}
\caption{The $\phi\left(s,\omega\right)$ diagram of $n=6$ TAE at $\omega_{{\rm A}0}t=700$
with only EP nonlinearity for $n_{{\rm EP}}/n_{{\rm e}}=0.002$
and $T_{{\rm EP}}=400{\rm keV}$. The black solid curves represent
the $n=6$ Alfv{\'e}n continuum shown in Figure \ref{fig:Density_q_profile}(a),
and the red solid curve is the radial envelope of TAE at the corresponding
time.}\label{fig:phi_s_omega_EP}
\end{figure}
\par\end{center}

\subsubsection{Nonlinear saturation with thermal plasma nonlinearity}\label{subsec:singlen_thermal}

With the nonlinear saturation of TAE due to EP nonlinearity investigated
above, the nonlinear saturation with both thermal plasma and EP nonlinearities
is investigated in this subsection. As illustrated in sec. \ref{sec:model},
the nonlinearity for each particle species can be turned on/off in
the code, so that we can compare the effects of each particle nonlinearity
on the nonlinear dynamics of TAE. The temporal evolution of the maximum
of $e\delta\phi_{n}/T_{\rm e}$ for cases given in Table \ref{tab:cases} is shown in Figure \ref{fig:phimax_no_zf_0.002}.
Comparing the cases A and D, it is found that the saturation level of TAE reduces significantly
from $e\delta\phi_{n}/T_{{\rm e}}\sim1$ in case D to $e\delta\phi_{n}/T_{{\rm e}}\sim0.1$ in case A due to  the introduction of thermal
plasma nonlinearities. To further validate the importance of thermal plasma nonlinearity in TAE saturation, the case
with nonlinear thermal species and linear EP is also given, i.e., case E. In this case, we can find that even in the absence of EP nonlinearity,
the mode can still saturate, and the saturation level is almost the
same as the case with nonlinearities of all particle species. In fact,
in the absence of EP nonlinearity, the mode should grow to infinity
regardless of the strength of the linear drive. However, according to the result given
here, the thermal plasma nonlinearity seems to impose a threshold
on TAE amplitude, so that the mode is saturated when its amplitude
reaches $e\delta\phi_{n}/T_{{\rm e}}\sim0.1$, even if there is sufficient
EP drive. Among nonlinearities of thermal plasmas, it is found that
the electron nonlinearity plays a dominant role in determining the
saturation level of TAE by comparing cases B and C, while the contribution
of thermal ions is modest. Thus, it is demonstrated that thermal plasma nonlinearity
plays a very important role in the nonlinear saturation of TAE, which
is not expected from the usual wave-particle trapping story. But this
has already been pointed out in previous literature, indicating the
importance of mode-mode coupling in nonlinear saturation of TAE \cite{LChenPPCF1998,FZoncaPRL1995}.

To identify the mechanism for TAE saturation by thermal plasma nonlinearity,
the evolution of TAE frequency at $s=0.5$
is shown in Figure \ref{fig:freq_no_zf_0.002}. It is found that, in the presence of thermal nonlinearities, the frequency of TAE decreases
at $\omega_{{\rm A0}}t\sim200$, while it is constant with only
EP nonlinearity. Thus, as the mode amplitude grows with time, the frequency shift induced by thermal plasma nonlinearity becomes more significant and leads to strong coupling with continuum, which may
finally result in the transition to energetic particle mode (EPM)
and saturation of TAE. This is proved by the $\phi\left(s,\omega\right)$
diagram shown in Figure \ref{fig:phi_s_omega_no_zf_all_nonlinearity}.
In the Figure \ref{fig:phi_s_omega_no_zf_all_nonlinearity}(a), there
is a TAE locates in TAE gap in the linear growth stage at $\omega_{{\rm A0}}t\simeq200$,
while, in the nonlinear saturation stage at $\omega_{{\rm A0}}t\simeq400$,
the $m=10$ and $m=11$ harmonics of TAE decouples and become two
independent EPMs chirping along Alfv{\'e}n continuum, as shown in Figure 
\ref{fig:phi_s_omega_no_zf_all_nonlinearity}(b). Meanwhile, as mentioned
before, the mode structure of TAE does not change with only EP nonlinearity.
Note here that due to the weak magnetic shear and low ion/electron
temperatures, there is little evidence of the generation of short wavelength
structure and parallel electric field associated with kinetic Alfv{\'e}n
wave (KAW) generation, as shown in Figure \ref{fig:E_parallel_ratio}.
The parallel electric field is obtained from $\delta E_{\parallel}=-\boldsymbol{b}\cdot\nabla\left(\delta\phi_{n}-\delta\psi_{n}\right)$,
and $\delta E_{\parallel,{\rm es}}=-\boldsymbol{b}\cdot\nabla\delta\phi_{n}$,
which is similar to what has been done in Ref. \cite{RMaarXiv2025}. It
is found that the ratio between $\delta E_{\parallel}$ and $\delta E_{\parallel,{\rm es}}$
is tripled from $\omega_{{\rm A0}}t=100$ to $400$, while $\delta\phi_{n}$
has increased by two orders of magnitude in this time interval. Thus, the excitation of a parallel electric field is not significant.

To have a thorough understanding of the nonlinear saturation of TAE, the scaling laws of saturation level versus linear drive for cases
A, D, and E are shown in Figure \ref{fig:scaling_law_no_zf}. For
simplicity, we can define three sets of saturation level $A_{{\rm sat}}^{{\rm ALL}}$,
$A_{{\rm sat}}^{{\rm EP}}$ and $A_{{\rm sat}}^{{\rm T}}$, representing
the saturation level of $e\delta\phi_{n}/T_{{\rm e}}$ with nonlinearities
of all particle species, only EP nonlinearity, only thermal plasma
nonlinearities, i.e., corresponding to cases A, D, and E.  It is found that $A_{{\rm sat}}^{{\rm T}}$ and $A_{{\rm sat}}^{{\rm ALL}}(\gamma_{{\rm L}}>\gamma_{{\rm L,C1}})$
has a very weak dependence on the linear drive, which indicates an
upper limit of the saturation level induced by thermal plasma nonlinearity, i.e., ``stiffness'' in saturation. Here, $\gamma_{{\rm L,C1}}$
is a critical linear drive given in Figure \ref{fig:scaling_law_no_zf},
which will be explained below. More specifically, as the mode amplitude
reaches $O(10^{-1})$, the mode saturates even if the free energy
in the phase space of EPs is not fully exhausted. As will be shown
in Equation (\ref{eq:thermal_sat_level}), as the amplitude of TAE
reaches $A_{{\rm sat}}^{{\rm T}}$, the nonlinear decoupling induced
by thermal plasma nonlinearity can essentially balance the linear
toroidal coupling of TAE, leading to the frequency downshift and
eventual saturation of TAE. Thus, defining the $\gamma_{{\rm L,C2}}$
as the linear drive satisfying $A_{{\rm sat}}^{{\rm EP}}=A_{{\rm sat}}^{{\rm T}}$,
we can have the boundary of low/high linear drive regime, which is
$\gamma_{{\rm L,C2}}\approx3200{\rm s}^{-1}$, or $\gamma_{{\rm L,C2}}/\omega_{0}\approx0.75\%$,
in the cases investigated here. In the regime with $\gamma_{{\rm L}}>\gamma_{{\rm L,C2}}$,
the nonlinear saturation of TAE is dominated by thermal plasma nonlinearity.
While, it is expected that, in the low linear drive regime with $\gamma_{{\rm L}}<\gamma_{{\rm L,C2}}$,
we should have $A_{{\rm sat}}^{{\rm ALL}}\simeq A_{{\rm sat}}^{{\rm EP}}$,
since EP nonlinearity should dominate over thermal plasma nonlinearity
in this regime. However, on the contrary, it is found that we have
$A_{{\rm sat}}^{{\rm ALL}}\simeq A_{{\rm sat}}^{{\rm T}}\sim0.05>A_{{\rm sat}}^{{\rm EP}}\simeq0.02$
when $\gamma_{{\rm L}}\approx2500{\rm s}^{-1}<\gamma_{{\rm L,C2}}$.
The time evolution of $e\delta\phi_n/T_{\rm e}$ for $\gamma_{{\rm L}}=2500{\rm s}^{-1}$
is shown in Figure \ref{fig:phimax_no_zf_0.0004}, in which one can observe
a longer linear growth stage in the presence of thermal plasma nonlinearities.
According to the evolution of mode frequency for corresponding cases shown in Figure \ref{fig:freq_no_zf_0004}, the possible explanation can be summarized in the Figure \ref{fig:schematic}.
The thermal plasma nonlinearity can induce a frequency downshift of
TAE via beat generation of $n=0$ PSZS as derived in sec. \ref{sec:theory},
which is proportionate to the intensity of TAE $|\delta\phi_{n}|^{2}$.
The frequency downshift can lead to a shift in the resonance region in the phase
space of EPs. Thus, even though the free energy near the original resonance
line is depleted, the mode can exhaust free energy in other parts
of EP phase space, which can lead to further growth of mode amplitude.
As a result, the thermal plasma nonlinearity can be enhanced, as can
the frequency downshift. This feedback loop enables a longer linear
growth stage until its amplitude reaches $A_{{\rm sat}}^{{\rm T}}$.
This mechanism is of little importance when $\gamma_{{\rm L}}>\gamma_{{\rm L,C2}}$,
since the free energy near the original resonance region is not fully
exhausted when the thermal plasma nonlinearity saturates the mode.
However, if $A_{{\rm sat}}^{{\rm EP}}$ is small enough that the frequency
downshift is below the threshold to trigger the feedback loop explained above, the nonlinear saturation of TAE should be dominated by EP
nonlinearity. Thus, we proceed further to a weaker linear drive, as
given by the leftmost point in Figure \ref{fig:scaling_law_no_zf}.
It is found that the saturation level with all nonlinearities is the
same with that determined by EP nonlinearity only, while the saturation
level with only thermal plasma nonlinearities is still given by the
threshold value determined by thermal plasma nonlinearity, i.e., $A_{{\rm sat}}^{{\rm T}}=0.04>A_{{\rm sat}}^{{\rm EP}}\sim A_{{\rm sat}}^{{\rm ALL}}=0.0049$.
Thus, there exists another critical linear drive value $\gamma_{{\rm L,C1}}=1300\sim2000{\rm s}^{-1}$, below which nonlinear saturation of TAE is dominated by EP nonlinearity; otherwise, it is dominated by thermal plasma nonlinearity. To determine
$\gamma_{{\rm L,C1}}$, the mode amplitude when the mode frequency
deviates from the linear frequency, e.g., it is $\omega_{{\rm A}0}t\simeq200$
for $n_{{\rm EP}}/n_{{\rm e}}=0.002$ shown in Figure \ref{fig:freq_no_zf_0.002},
is studied, which is $e\delta\phi_{n}/T_{{\rm e}}\approx0.01$ for
all the linear drives given in Figure \ref{fig:scaling_law_no_zf}.
This implies that $\gamma_{{\rm L,C1}}$ is the linear drive corresponds
to $A_{{\rm sat}}^{{\rm EP}}\left(\gamma_{{\rm L,C1}}\right)\simeq0.01$,
which is close to $\gamma_{{\rm L}}=2000{\rm s^{-1}}$. In summary, the saturation level of TAE in a single-$n$ simulation can be given
by 
\begin{eqnarray*}
 &  & A_{{\rm sat}}^{{\rm EP}},\gamma_{{\rm L}}<\gamma_{{\rm L,C1}}\\
A_{{\rm sat}} & = & A_{{\rm sat}}^{{\rm T}},\gamma_{{\rm L,C2}}>\gamma_{{\rm L}}>\gamma_{{\rm L,C1}}\\
 &  & A_{{\rm sat}}^{{\rm T}},\gamma_{{\rm L}}>\gamma_{{\rm L,C2}}
\end{eqnarray*}

The reason to present them in three intervals instead of two is that the physical mechanism behind the second and third intervals is different, though the saturation level of TAE is the same. The theoretical analysis
on the shifting of resonance region and determination of $\gamma_{{\rm L,C1}}$
are beyond the scope intended in this paper and might be analyzed
in future publications.

To give guidance to the theoretical derivation, the effect of parallel
nonlinearity is investigated by imposing $\dot{\delta v_{\parallel}}=0$
on the left-hand-side of Equation (\ref{eq:gk_vlasov}). It is found that the temporal evolution and saturation level of TAE are not affected in the absence of the parallel nonlinearity from the ORB5 simulation.

This can be explained theoretically. The parallel nonlinearity is given by $\delta\dot{v}_{\parallel}\partial\delta f_{{\rm s}}/\partial v_{\parallel}$,
which can be readily obtained as
\begin{eqnarray*}
\delta\dot{v}_{\parallel}\dfrac{\partial\delta f_{{\rm s}}}{\partial v_{\parallel}} & = & \dfrac{q_{{\rm s}}}{m_{{\rm s}}}\left(\delta E_{\parallel n}\dfrac{\partial\delta f_{{\rm s,Z}}}{\partial v_{\parallel}}+\delta E_{\parallel{\rm Z}}\dfrac{\partial\delta f_{{\rm s},n}}{\partial v_{\parallel}}\right).
\end{eqnarray*}

The first term vanishes because of the ideal MHD constraint with $\delta\phi\approx\delta\psi$ is still satisfied near mode saturation, which is
shown in Figure \ref{fig:E_parallel_ratio}. For both cases with and without zonal fields, the $\delta\phi\approx\delta\psi$
is satisfied near saturation. The second term is also of little importance
since the particle responses of TAE do not have anti-symmetric dependence
on $v_{\parallel}$. 
\begin{center}
\begin{figure}
\begin{centering}
\includegraphics[scale=0.4]{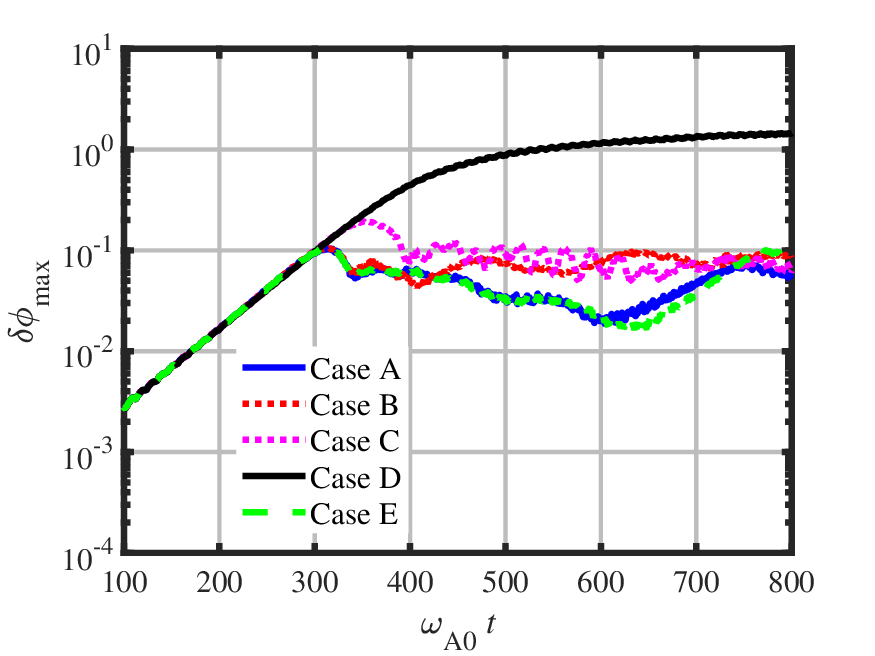}
\par\end{centering}
\caption{The temporal evolution of the maximum of electrostatic potential 
$e\delta\phi_n/T_{\rm e}$
of TAE in single-$n$ simulations for five cases shown in the table \ref{tab:cases}, i.e., case A: with all nonlinearities, case B: with only electron nonlinearity, case C: with only ion nonlinearity, case D: with only EP nonlinearity, and case E: with only thermal plasma nonlinearities.}\label{fig:phimax_no_zf_0.002}
\end{figure}
\par\end{center}

\begin{center}
\begin{figure}
\includegraphics[scale=0.4]{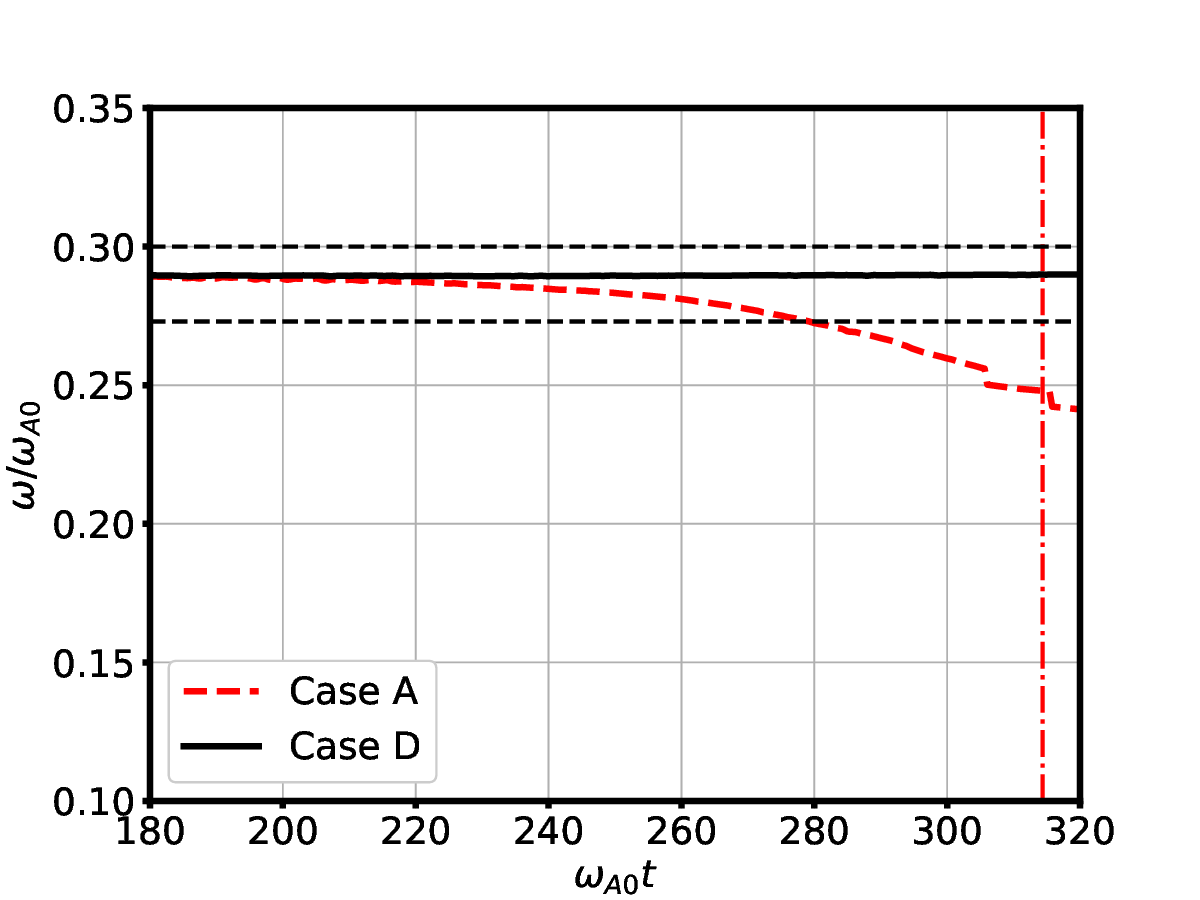}

\caption{The temporal evolution of the frequency of TAE at $s=0.5$, with the
red dashed curve representing the case with all nonlinearities, i.e., case A, and the blue solid line represents the case with only EP nonlinearity, i.e., case D.
The black dashed horizontal lines represent the upper and lower continuum,
respectively. The red dashed vertical line represents the time
for case A to saturate. }\label{fig:freq_no_zf_0.002}

\end{figure}
\par\end{center}

\begin{center}
\begin{figure}
\begin{centering}
\subfloat{\includegraphics[scale=0.18]{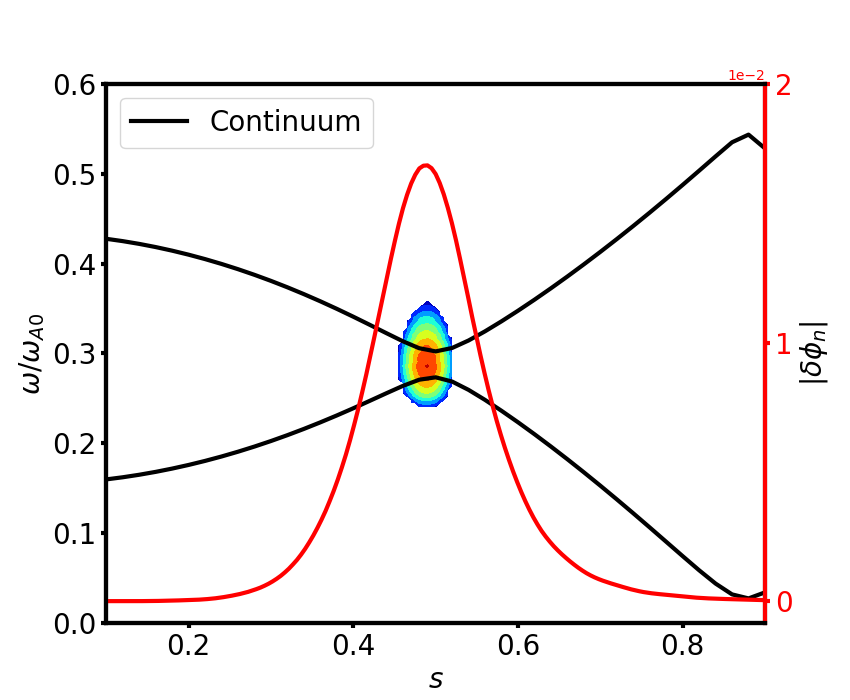}}\ \subfloat{\includegraphics[scale=0.265]{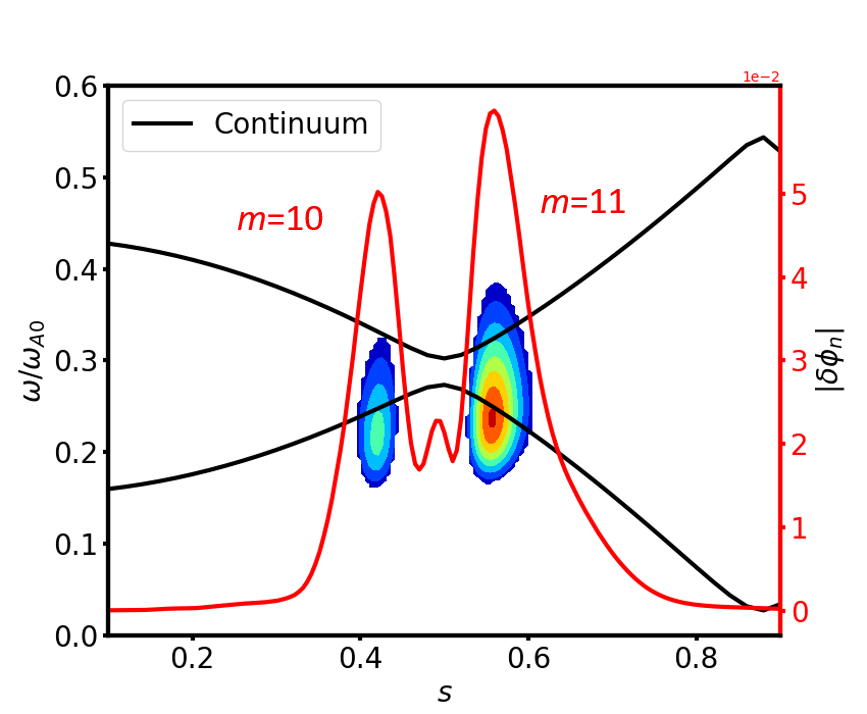}}
\par\end{centering}
\caption{The $\phi\left(s,\omega\right)$ diagram of $n=6$ TAE at (a) $\omega_{{\rm A}0}t=200$,
and (b) $\omega_{{\rm A}0}t=400$ for the simulation with all nonlinearities. The black solid curves represent
the $n=6$ Alfv{\'e}n continuum shown in Figure \ref{fig:Density_q_profile}(a),
and the red solid curve is the radial mode structure of TAE at the corresponding
time. }\label{fig:phi_s_omega_no_zf_all_nonlinearity}
\end{figure}
\par\end{center}

\begin{center}
\begin{figure}
\includegraphics[scale=0.5]{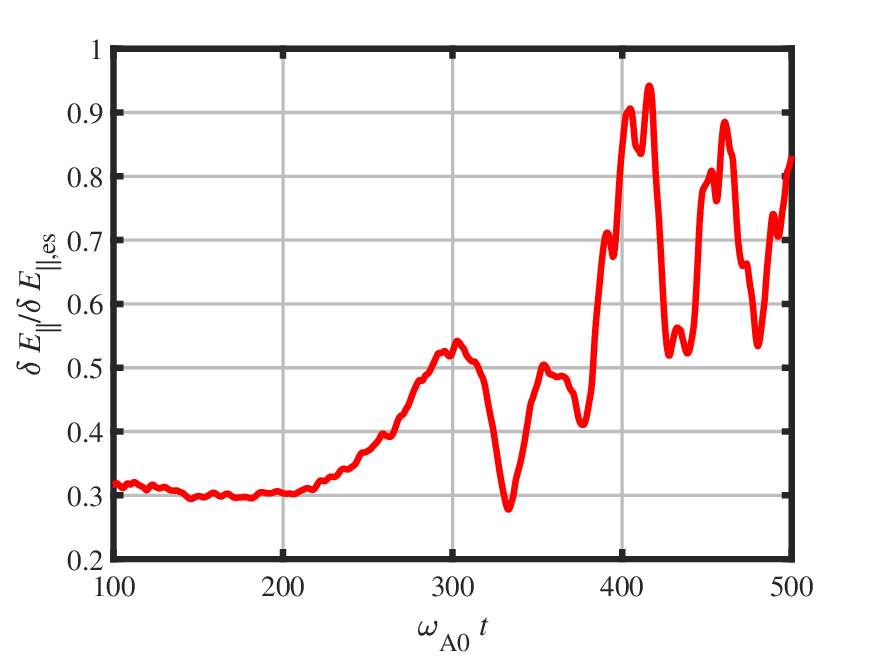}

\caption{The temporal evolution of the ratio between the parallel electric field $\delta E_{\parallel}$ and electrostatic parallel electric field $\delta E_{\rm \parallel,es}$, with $\delta E_{\parallel}\equiv-\boldsymbol{b}\cdot\nabla(\delta\phi_n-\delta\psi_n)$ and $\delta E_{\rm \parallel,es}\equiv-\boldsymbol{b}\cdot\nabla\delta\phi_n$.
Note that the parameter used in the rightmost
case is $n_{{\rm EP}}/n_{{\rm e}}=0.002$, $T_{{\rm EP}}=400{\rm keV}$.}\label{fig:E_parallel_ratio}
\end{figure}
\par\end{center}

\begin{center}
\begin{figure}
\begin{centering}
\includegraphics[scale=0.4]{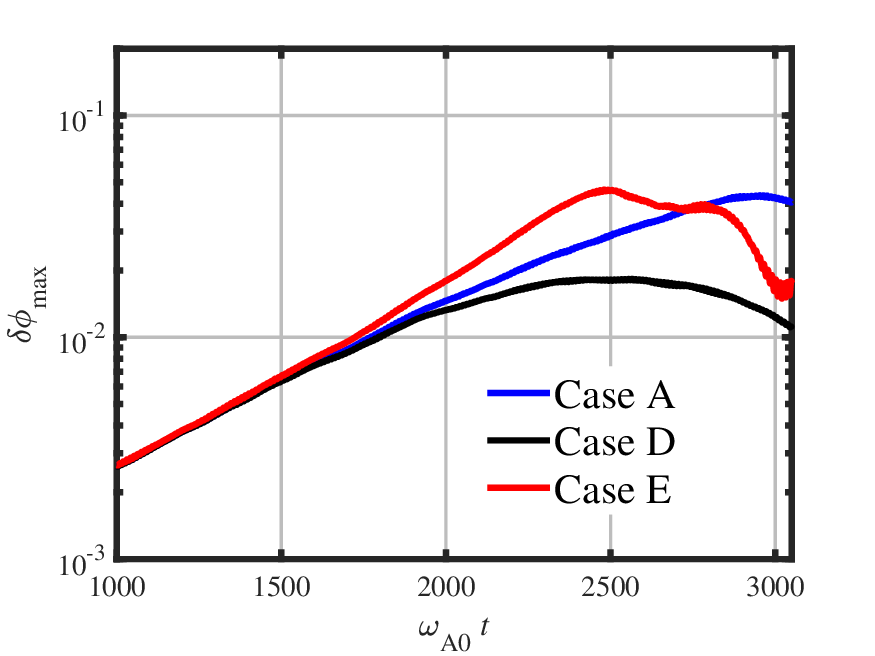}
\par\end{centering}
\caption{The temporal evolution of the maximum of electrostatic potential $\delta\phi$
of TAE for cases A: with all nonlinearities, D: with only EP nonlinearity and E: with only thermal plasma nonlinearities, which are indicated with blue, black
and red solid curves. Here, $n_{{\rm EP}}/n_{{\rm e}}=0.0004$, which corresponds to the second point from the left in Figure \ref{fig:scaling_law_no_zf}.}\label{fig:phimax_no_zf_0.0004}
\end{figure}
\par\end{center}

\begin{center}
\begin{figure}
\begin{centering}
\includegraphics[scale=0.4]{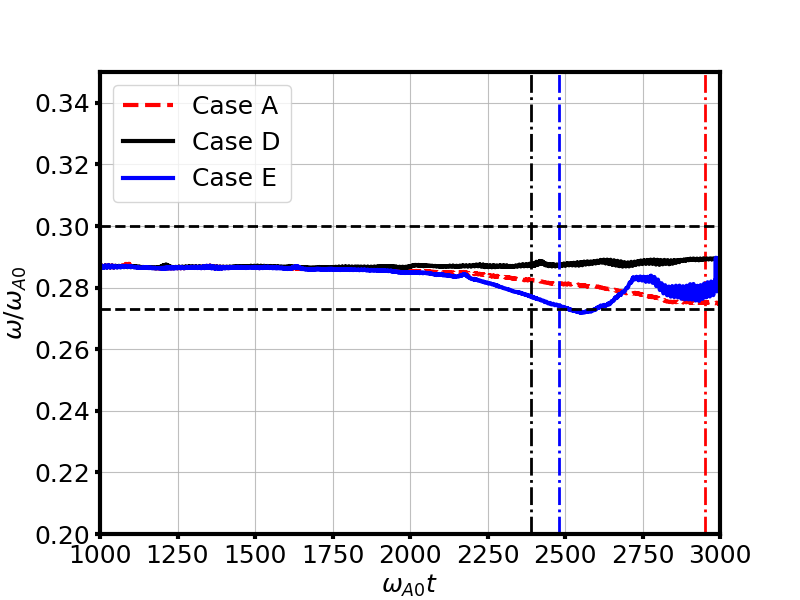}
\par\end{centering}
\caption{The temporal evolution of the frequency of TAE at $s=0.5$ corresponding to the case in Figure. \ref{fig:phimax_no_zf_0.0004}, with the
black and blue solid curves representing the cases D: with only EP nonlinearity and E: with only thermal plasma nonlinearities, and the red dashed curve indicates case A: with all nonlinearities. The black dashed horizontal lines represent the upper and lower continuum, respectively. The black and blue dashed vertical lines represent the time for the
corresponding case to saturate.}\label{fig:freq_no_zf_0004}
\end{figure}
\par\end{center}

\begin{center}
\begin{figure}
\begin{centering}
\includegraphics[scale=0.4]{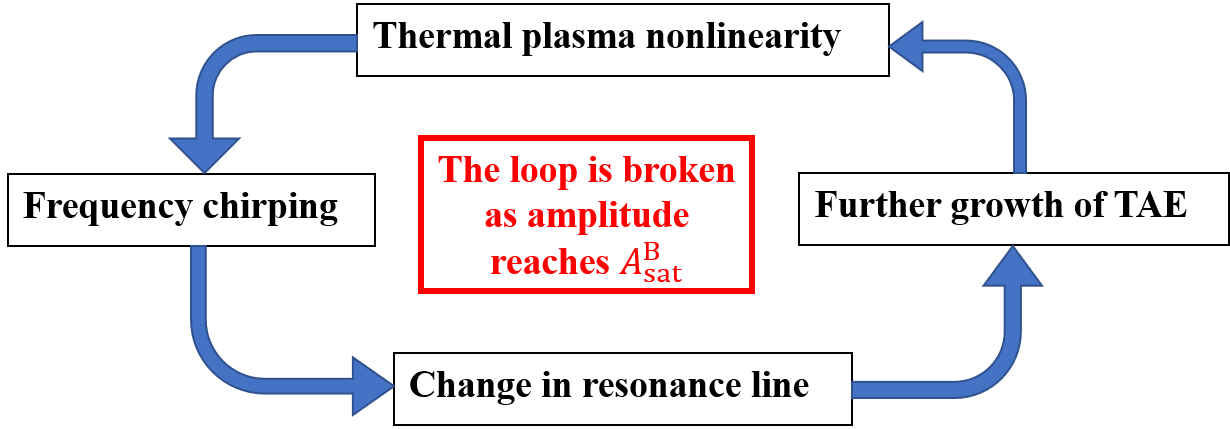}
\par\end{centering}
\caption{The cartoon showing the feedback loop of nonlinear growth and saturation
of TAE for linear drive in the range $\gamma_{{\rm L,C2}}>\gamma_{{\rm L}}>\gamma_{{\rm L,C1}}$.
}\label{fig:schematic}
\end{figure}
\par\end{center}

\subsection{Simulation results with zonal fields}\label{subsec:with_ZS}

In the previous section, the effects of thermal plasma and EP nonlinearities
on TAE saturation is investigated in single-$n$ simulations with $n=0$ zonal fields filtered out, while PSZS of thermal species exists when thermal plasma nonlinearities are retained. In fact, PSZS and zonal fields are generated simultaneously; thus, the imbalanced system investigated above may over-/under-estimate the saturation level of TAE. Then, the $n=0$ modes should
be retained to get a more self-consistent result. In this section,
the nonlinear saturation of TAE will be investigated by retaining
both $n=6$ TAE and $n=0$ zonal fields. Note again that, according to the convergence
study shown in the appendix. A, the realistic mass ratio is used
in this case. Meanwhile, only the cases with the nonlinearities
of all particle species are studied here, i.e., case A, due to the
fact that zonal fields excitation is related to the polarization
density perturbation, or the system is not self-consistent. 

The temporal evolution of the maximum of $e\delta\phi_{n}/T_{\rm e}$ and zonal
electric field $\partial_{s}\delta\phi_{{\rm Z}}$ are shown in the
Figure \ref{fig:with_without_ZF_compare_phimax}. Clear evidence
of a doubled growth rate is observed, indicating the beat-driven of zonal
fields by TAE. Meanwhile, comparing the blue and black curves, we found
that the saturation level of TAE changes from $A_{{\rm sat}}^{{\rm ALL}}\simeq0.1$
for the single-$n$ simulation to $A_{{\rm sat}}^{{\rm Z}}\sim0.26$,
where $A_{{\rm sat}}^{{\rm Z}}$ is the saturation level of TAE with
zonal fields retained. This result is dramatically different from
common speculation that beat-driven zonal fields can stabilize TAE. Note that here we are not to indicate a more unstable TAE with higher saturation level due to zonal fields; the enhancement here is because PSZS always exists in PIC codes once the corresponding particle species evolves nonlinearly. For MHD codes, the PSZS and zonal fields can be turned on/off freely, then the effects of zonal fields might be identified solely in the MHD limit, which was found to be stabilizing \cite{YTodoNF2010}. To examine this phenomenon in a broader parameter range, the scaling law of the saturation level of TAE in the cases with and without $n=0$ modes is shown in Figure \ref{fig:With_ZF_sat_gamma_scaling}. We can observe that for reasonable linear growth rate $\gamma_{\rm L}/\omega_n>1\%$, the saturation level with $n=0$ modes is roughly twice as large as that in a single-$n$ simulation. Meanwhile, the ``stiffness'' in saturation can also be found in the case with zonal fields, with a saturation level being $e\delta\phi_n/T_{\rm e}\sim0.2$, or $\delta B_r/B_0\sim10^{-4}$. The results given here imply that for future tokamaks with stronger EP drive, the ``stiffness'' in saturation of TAE induced by the thermal plasma nonlinearities might result in a comparable TAE level to current devices with lower EP drive.

Recall that the beat-driven PSZS of thermal species can lead to a reduction of mode frequency
and early saturation, the enhancement of saturation level of TAE indicates
that the presence of zonal fields might counteract the effect
of PSZS and the consequent frequency downshift. The evolution of mode frequency is studied and shown in Figure \ref{fig:with_without_ZF_freq_0.002}.
It is found that
the decrease of mode frequency happens at $\omega_{{\rm A}0}t\sim330$
for the case with zonal fields, while it is $\omega_{{\rm A}0}t\sim200$
for the case without zonal fields. This indicates that the amplitude of TAE needed
for the mode frequency to decrease becomes larger when $n=0$ modes
are retained, which is consistent with statements above. 

The analysis given above focuses on the evolution of TAE, the mechanism
for zonal fields generation should also be investigated. There are mainly two
mechanisms responsible for beat-driven of zonal fields, i.e., beat-driven due
to thermal plasma nonlinearities and resonant EP contribution. The first
channel was firstly investigated in Ref. \cite{LChenNF2025} for the reversed shear
Alfv{\'e}n eigenmode (RSAE), and that for TAE was derived in Ref. \cite{QFangNF2025}. Following
the same procedure, the result is reproduced in Equation (\ref{eq:thermal_ZF}).
Meanwhile, zonal fields generated by resonance EP contribution is
given in equations. (23) and (31) in Ref. \cite{ZQiuNF2017}. Here, we define $\delta\phi_{{\rm Z}}^{{\rm T}}$
and $\delta\phi_{{\rm Z}}^{{\rm E}}$ to represent zonal scalar potential
generated by thermal plasma nonlinearity and resonant EPs contribution,
respectively. The quantitative comparison of the radial mode structure
of zonal fields are made between the simulation and analytical results,
which are shown in Figure \ref{fig:radial_structure_ZF_ZC}(a) and
(b) for zonal scalar potential and zonal current, respectively. It
is found that the simulation result of $\partial_{s}\delta\phi_{{\rm Z}}$
is about $250$ times larger than that due to thermal plasma $\partial_{s}\delta\phi_{{\rm Z}}^{{\rm T}}$,
i.e., $\partial_{s}\delta\phi_{{\rm Z}}\approx250\partial_{s}\delta\phi_{{\rm Z}}^{{\rm T}}$,
and $\partial_s\delta\phi_{{\rm Z}}^{{\rm E}}$ agrees well with simulation
results. Thus, the resonant EP contribution dominates over thermal
plasma nonlinearity in zonal fields generation, due to the flat thermal plasma
profile and low-$n$ TAE in this case. Note that we have used $k_{\perp}^{2}\approx\partial_{r}^{2}/2$
in obtaining $\partial_{s}\delta\phi_{{\rm Z}}^{{\rm E}}$, since
the simulation result is clearly a second-order derivative of Gaussian
envelope. The inclusion of $k_{\theta}^{2}$ can lead to distortion
of the mode structure, though the amplitude is similar. A good agreement
of the zonal vector potential $\delta A_{\parallel,{\rm Z}}$ can
also be obtained, as shown in Figure \ref{fig:radial_structure_ZF_ZC}(b). The analytical result $\partial_s\delta A_{\rm Z,a}$ is obtained from Equation \ref{eq:AZ}. More evidence on the little contribution of thermal plasma nonlinearity to zonal flow excitation is shown in Figure \ref{fig:radial_structure_nonunifomity}.
By adjusting the characteristic length of thermal ion non-uniformity
$L_{{\rm ni}}\equiv -n_{\rm i}/({\rm d}n_{\rm i}/{\rm d}r)$, the radial mode structure of zonal electric field
at $\omega_{{\rm A}0}t=320$ is given in Figure \ref{fig:radial_structure_nonunifomity}(a).
In fact, the growth rate of TAE is dependent on $L_{{\rm ni}}$, so
that the amplitude of TAE is different for different $L_{n{\rm i}}$.
With the normalization $\partial_{s}\delta\phi_{{\rm Z}}/|\delta\phi_{n}|^{2}$
used, the quantity proportionate to the nonlinear coupling coefficient can be obtained, as shown in Figure \ref{fig:radial_structure_nonunifomity}(b).
For the zonal scalar potential generated by thermal plasma nonlinearity,
$\partial_{s}\delta\phi_{{\rm Z}}/|\delta\phi_{n}|^{2}$ should be
inversely proportionate to $L_{{\rm ni}}$, as shown by Equation 
(\ref{eq:thermal_ZF}), but it is almost the same for four $L_{{\rm ni}}$
values investigated here. Thus, we have $\delta\phi_{{\rm Z}}\approx\delta\phi_{{\rm Z}}^{{\rm E}}\approx250\delta\phi_{{\rm Z}}^{{\rm T}}$,
which will be used in the theoretical interpretation given below.

\begin{center}
\begin{figure}
\begin{centering}
\includegraphics[scale=0.4]{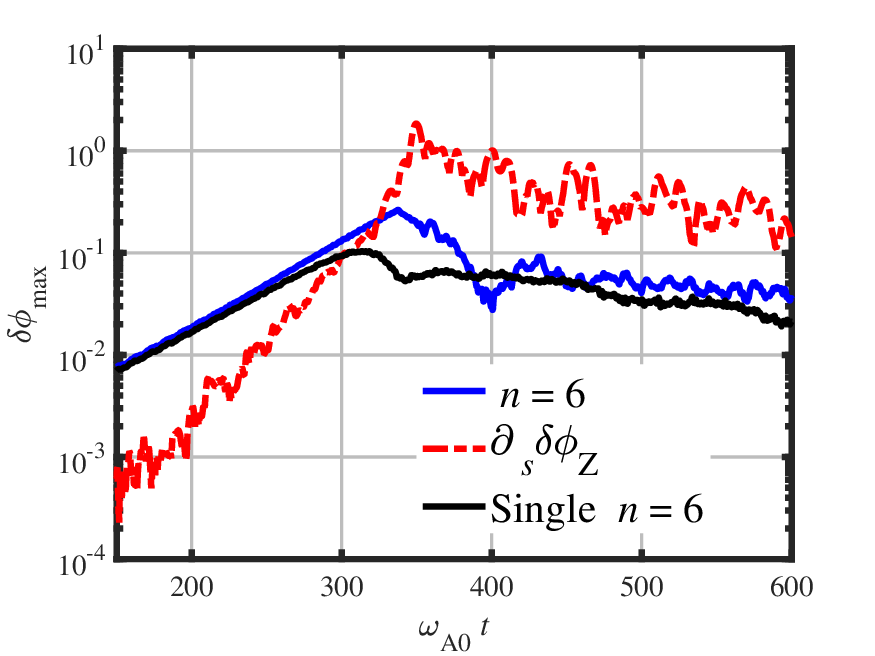}
\par\end{centering}
\caption{The temporal evolution of the maximum of electrostatic potential $\delta\phi_n$
of TAE and radial electric field of $n=0$ zonal fluctuations are shown by the blue solid and red dashed lines, respectively. Meanwhile, the result of
$\delta\phi_n$ in single-$n$ simulation is given in black solid curve for comparison.}\label{fig:with_without_ZF_compare_phimax}
\end{figure}
\par\end{center}

\begin{center}
\begin{figure}
\begin{centering}
\includegraphics[scale=0.4]{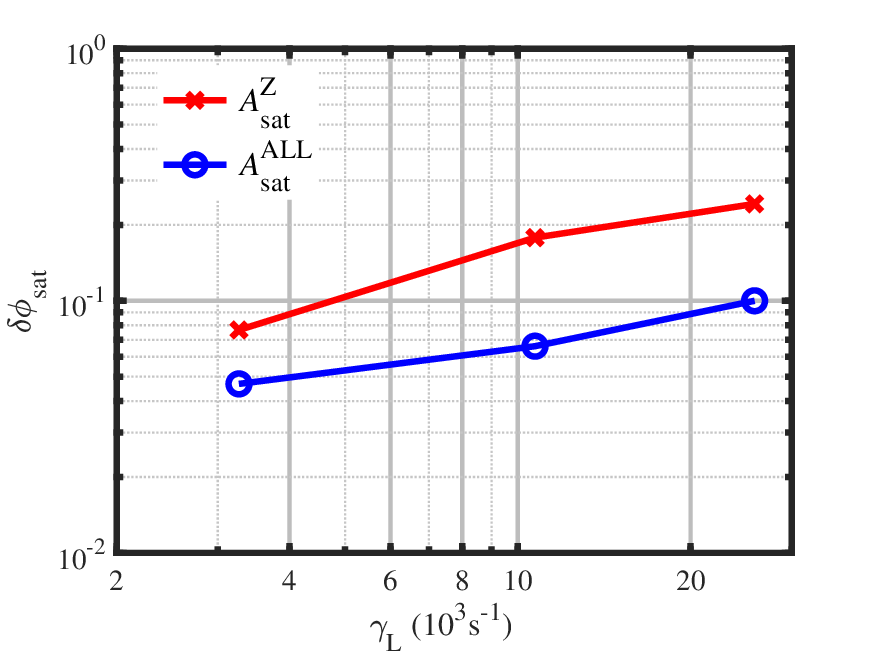}
\par\end{centering}
\caption{The dependence of saturation level of TAE on the linear drive for the case with zonal fields $A_{\rm sat}^{\rm Z}$ (red crosses) and without zonal fields $A_{\rm sat}^{\rm ALL}$(blue circles). The blue line corresponds to the  first three points from the right of Figure \ref{fig:scaling_law_no_zf}.}\label{fig:With_ZF_sat_gamma_scaling}
\end{figure}
\par\end{center}

\begin{center}
\begin{figure}
\begin{centering}
\includegraphics[scale=0.4]{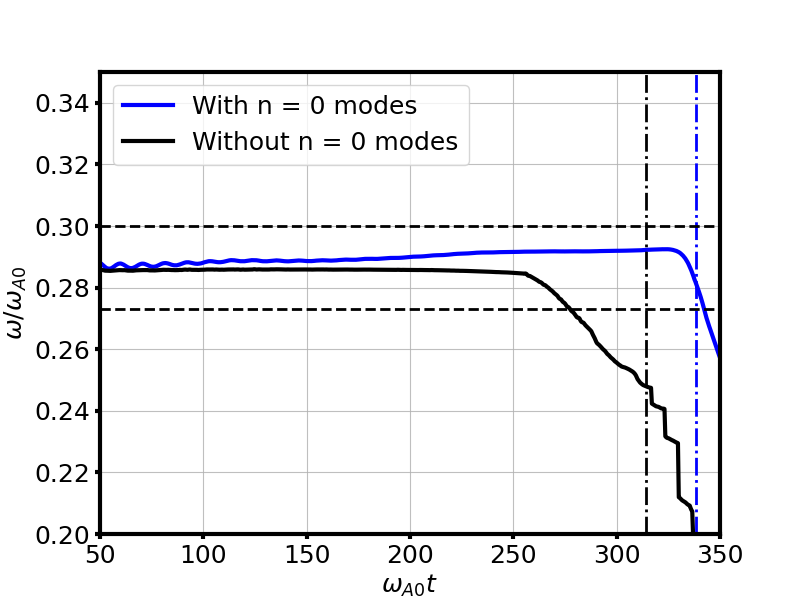}
\par\end{centering}
\caption{The temporal evolution of the frequency of TAE at $s=0.5$, with the
blue and black lines representing the case with $n=0$ modes and
the single-$n$ simulation. Again, the black dashed horizontal lines
represent the upper and lower continuum, respectively. Meanwhile, the
black and blue dashed vertical lines represent the time for the corresponding
case to saturate.}\label{fig:with_without_ZF_freq_0.002}
\end{figure}
\par\end{center}

\begin{center}
\begin{figure}
\begin{centering}
\subfloat{\includegraphics[scale=0.26]{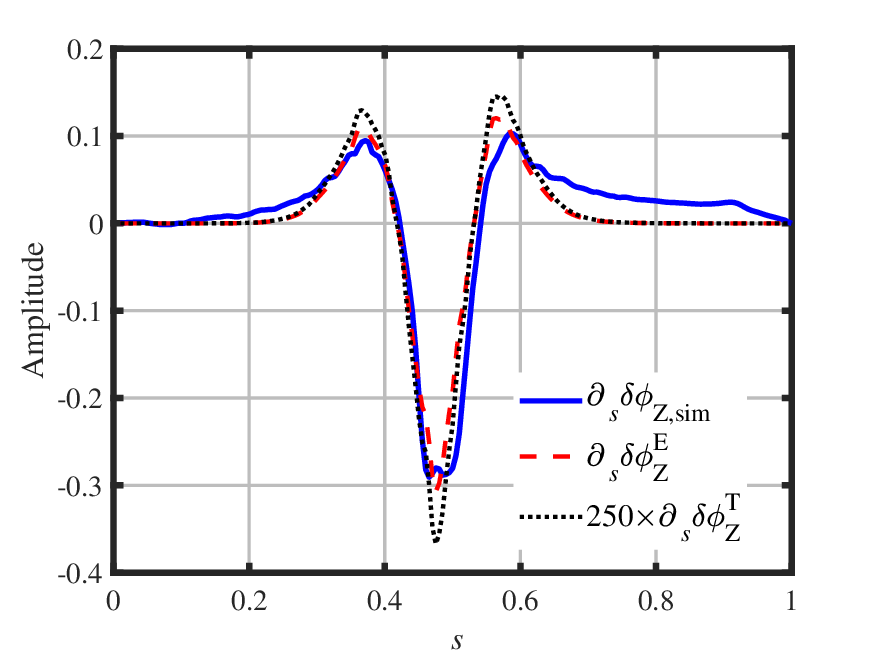}}\ \subfloat{\includegraphics[scale=0.26]{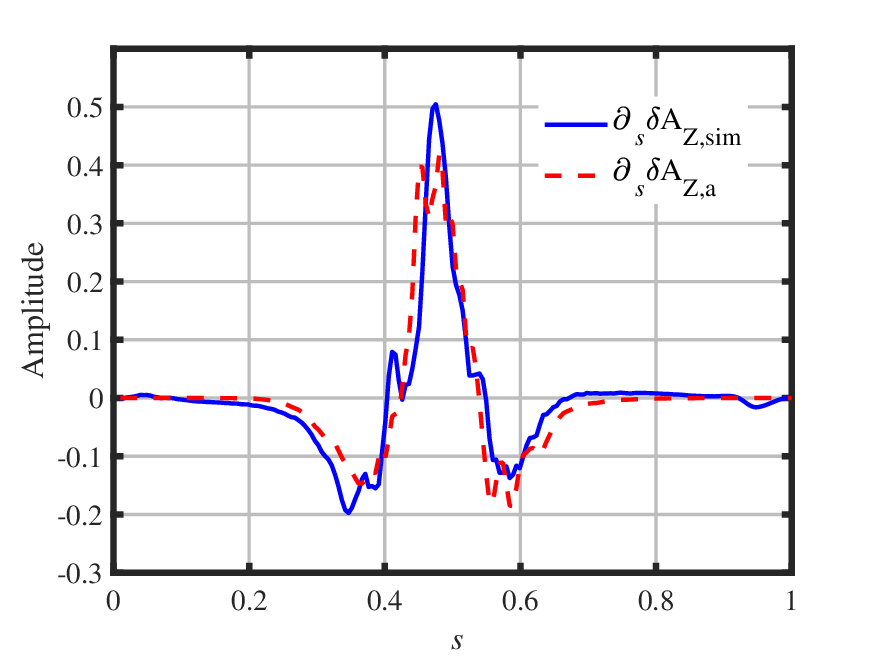}}
\par\end{centering}
\caption{The radial mode structure of (a) radial electric field of zonal fields, i.e.,
$\partial_{s}\delta\phi_{{\rm Z}}$, and (b) radial derivative of
zonal vector potential for $n_{{\rm EP}}/n_{{\rm e}}=0.002$,
and $T_{{\rm EP}}=400{\rm keV}$. $\delta\phi_{{\rm Z}}^{{\rm T}}$
and $\delta\phi_{{\rm Z}}^{{\rm E}}$ denote the theoretical zonal scalar potential 
generated by thermal plasma nonlinearity given by Equation (\ref{eq:thermal_ZF}) and resonant EPs contribution given by Equation (31) in Ref. \cite{ZQiuNF2017},
respectively. The subscripts ``sim'' and ``a''
represent that obtained from the simulation and analytical expression,
respectively. }\label{fig:radial_structure_ZF_ZC}
\end{figure}
\par\end{center}

\begin{center}
\begin{figure}
\begin{centering}
\subfloat{\includegraphics[scale=0.26]{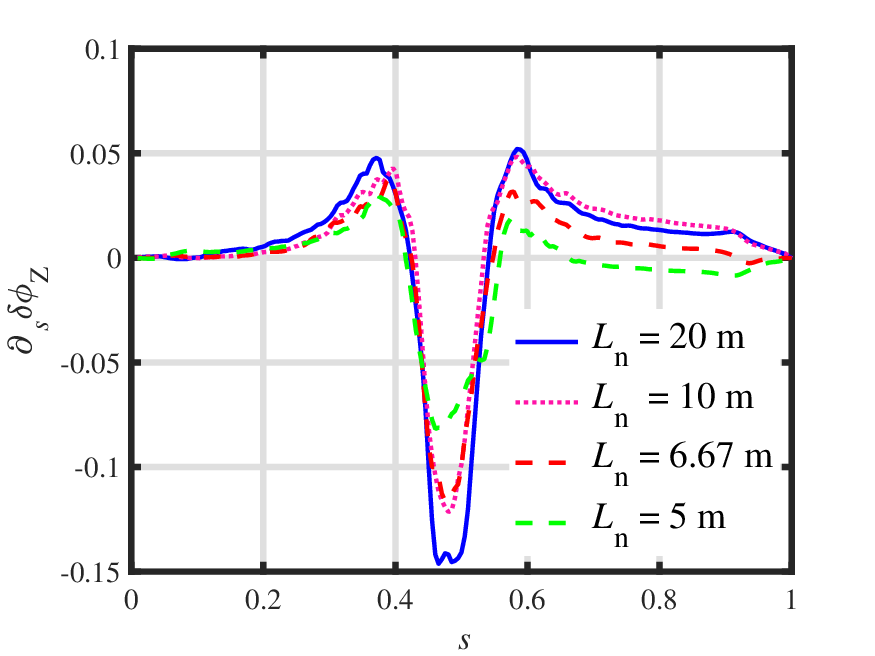}}\ \subfloat{\includegraphics[scale=0.26]{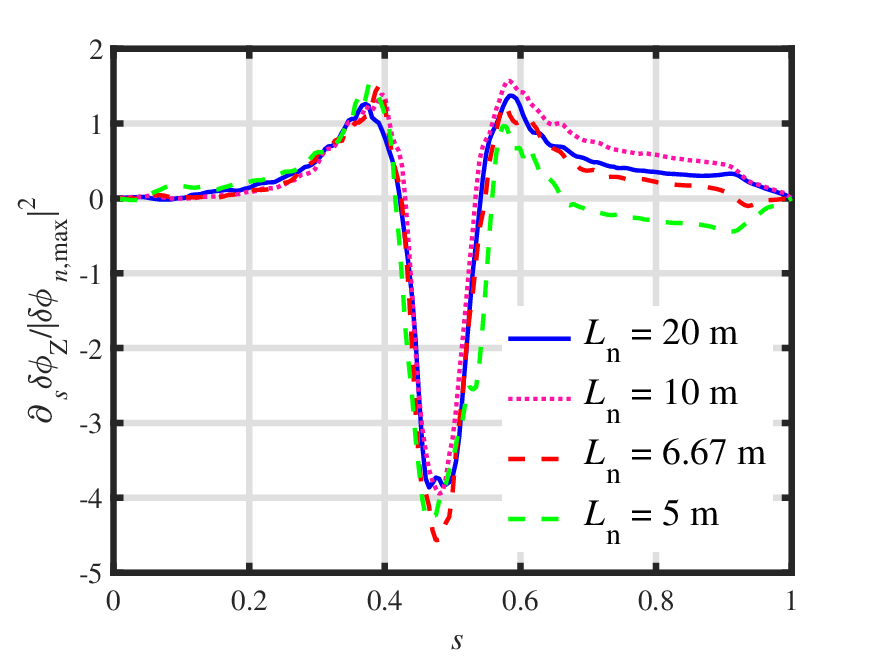}}
\par\end{centering}
\caption{The radial mode structure of (a) radial electric field of zonal fields, i.e.,
$\partial_{s}\delta\phi_{{\rm Z}}$, and (b) modified radial electric
field normalized by maximum of TAE intensity, i.e., $\partial_{s}\delta\phi_{{\rm Z}}/\left|\delta\phi_{n}\right|^{2}$
for four values of $L_{{\rm ni}}$. }\label{fig:radial_structure_nonunifomity}
\end{figure}
\par\end{center}

\section{Theoretical interpretation}\label{sec:theory}

In this section, a theoretical model based on the nonlinear gyrokinetic
framework is developed to explain the numerical results shown in the previous
section. The nonlinear feedback of PSZS and zonal fields to TAE is
derived in the first place. Then, to make a correspondence to simulations
in the previous section, the saturation is analyzed with and
without $n=0$ zonal fields, while both of them include PSZS of thermal species.

For single-$n$ TAE with $\Omega_{n}=\left(\omega_{n},k_{\theta n}\right)$, it can couple with its complex conjugate to beat-driven zonal fields with $\Omega_{{\rm Z}}=\left(\omega_{{\rm Z}},k_{r{\rm Z}}\right)$
and also $n=0$ PSZS \cite{YTodoNF2010}. To focus on the effects of thermal plasma nonlinearities, the EP nonlinearity is neglected. Recall that parallel nonlinearity has little effect on TAE saturation, thermal plasma responses
to TAE and zonal fields can be derived from the nonlinear gyrokinetic
equation with convective perpendicular nonlinearity and quasi-neutrality
condition \cite{EFriemanPoF1982}. The nonlinear nonadiabatic particle responses to zonal
fields, i.e., PSZS, for thermal plasmas in the drift-kinetic limit can
be derived as
\begin{eqnarray}
\overline{\delta H_{{\rm Z,i}}^{{\rm NL}}} & = & -\dfrac{c}{B_{0}}\dfrac{e}{T_{{\rm i}}}\dfrac{\omega_{*{\rm i}}}{\omega_n^{2}}F_{0{\rm i}}\partial_{r}\left(k_{\theta n}\left|\delta\phi_{n}\right|^{2}\right),\label{eq:Zonal_NL_ion}
\end{eqnarray}
\begin{eqnarray}
\delta H_{{\rm Z,{\rm e}}}^{{\rm NL}} & = & \dfrac{e}{T_{{\rm e}}}\dfrac{c}{B_{0}}F_{0{\rm e}}\partial_{r}k_{\theta n}\left(\dfrac{\omega_{*{\rm e},n}}{\omega_{n}^{2}}-\dfrac{k_{\parallel}v_{\parallel}}{\omega_{n}^{2}}\right)\left|\delta\phi_{n}\right|^{2}.\label{eq:Zonal_NL_electron}
\end{eqnarray}

Here, the nonadiabatic particle responses $\delta H_{k,{\rm s}}$
can be separated into linear and nonlinear components, i.e., $\delta H_{k,{\rm s}}\equiv\delta H_{k,{\rm s}}^{{\rm L}}+\delta H_{k,{\rm s}}^{{\rm NL}}$,
with the subscript ``Z'' and ``$n$'' representing quantities associated
with zonal fields and TAE, $\omega_{*{\rm s}}\equiv k_{\theta}cT_{{\rm s}}/(eB_{0}L_{n})$,
$L_{n{\rm s}}\equiv-n_{\rm s}/({\rm d}n_{\rm s}/{\rm d}r)$ is the characteristic
length of equilibrium density profile, $k_{\theta}=|n|q/r$ is the
poloidal mode number, $e$ is the unit charge, $F_{0{\rm s}}$ is the
equilibrium distribution function of species ``$s$'', where Maxwellian distribution is adopted for thermal plasmas. Other notations are standard. It is
worthwhile noting that the second term in the PSZS of electron, i.e.,
Equation (\ref{eq:Zonal_NL_electron}), has an odd dependence on $v_{\parallel}$,
which essentially corresponds to the perturbed parallel current. The
PSZS of thermal ions only contributes to the density perturbation. Substituting
linear and nonlinear particle responses to the quasineutrality condition, the beat-driven zonal scalar potential due to thermal plasma nonlinearities can be obtained as
\begin{eqnarray}
\delta\phi_{{\rm Z}}^{{\rm T}} & = & \dfrac{c}{B_{0}}\dfrac{1}{\omega_{n}^{2}}\partial_{r}\sum_{m}\omega_{*{\rm i}}k_{\theta n}\left|\delta\phi_{m}\right|^{2}.\label{eq:thermal_ZF}
\end{eqnarray}

Meanwhile, the perturbed vector potential $\delta A_{\parallel,{\rm Z}}$
can be obtained from the parallel Ampere's law $\left(c^{2}/(4\pi)\right)\nabla_{\perp}^{2}\delta A_{\parallel,{\rm Z}}=\delta j_{{\rm Z}}$,
where $\delta j_{{\rm Z}}$ is the perturbed parallel current corresponding
to PSZS of electron. Then, the perturbed vector potential can be obtained
as 
\begin{eqnarray}
\delta A_{\parallel,{\rm Z}} & = & \dfrac{c^{2}}{B_{0}}\dfrac{1}{\omega_{n}^{2}}\partial_{r}\sum_{m}k_{\theta n}k_{\parallel}\left|\delta\phi_{m}\right|^{2}.\label{eq:AZ}
\end{eqnarray}

Then, the results in Ref. \cite{ZQiuNF2017,LChenNF2025,QFangNF2025} are reproduced. The nonlinear feedback
of zonal fields and PSZS to TAE can also be obtained as 
\begin{eqnarray}
\delta H_{n,{\rm {\rm i}}}^{{\rm NL}} & = & \dfrac{c}{B_{0}\omega_{n}}k_{\theta n}\dfrac{e}{T_{{\rm i}}}F_{0{\rm i}}\left(\dfrac{\omega_{*{\rm i}}}{\omega}\delta\phi_{n}\partial_{r}\delta\phi_{{\rm Z}}\right.\nonumber \\
 &  & \left.+\dfrac{k_{\parallel}v_{\parallel}^{2}}{\omega_{n}c}\delta\phi_{n}\partial_{r}\delta A_{\parallel,{\rm Z}}-\delta\phi_{n}\partial_{r}\delta\phi_{{\rm Z}}^{{\rm T}}\right),\label{eq:TAE_ion_NL_full}
\end{eqnarray}

\begin{eqnarray}
\delta H_{n,{\rm e}}^{{\rm NL}} & = & \dfrac{e}{T_{{\rm e}}}\dfrac{c}{B_{0}\omega_{n}}k_{\theta n}F_{0{\rm e}}\nonumber \\
 &  & \times\left(-\delta\phi_{n}\partial_{r}\delta\phi_{{\rm Z}}+\dfrac{\omega_{n}}{\omega_{*{\rm i}}}\delta\phi_{n}\partial_{r}\delta\phi_{{\rm Z}}^{{\rm T}}\right).\label{eq:TAE_electron_NL_full}
\end{eqnarray}

Note that the term proportionate to $\omega_{n}/\omega_{*{\rm i}}$
in the equation (\ref{eq:TAE_electron_NL_full}) originates from the
term with anti-symmetric $v_{\parallel}$ dependence, i.e., parallel
current beat driven by TAE. In both thermal ion and electron responses,
the last term are the contribution of PSZS, while other terms originate
from zonal fields. The nonlinear feedback of zonal fields and PSZS
to TAE enters through the curvature coupling term (CCT) in the nonlinear
vorticity equation
\begin{eqnarray}
\dfrac{c^{2}B_{0}}{4\pi\omega^{2}}\partial_{l}\left(\dfrac{k_{\perp}^{2}}{B_{0}}\right)\partial_{l}\delta\psi_{k}+\dfrac{e^{2}}{T_{{\rm i}}}\left\langle \left(1-{\rm J}_{0}^{2}\right)F_{0{\rm s}}\right\rangle \delta\phi_{k}\nonumber \\
-\sum_{{\rm s}}\left\langle \dfrac{q_{{\rm s}}}{\omega}\omega_{{\rm Ds}}\delta H_{{\rm s}}\right\rangle _{k}=0,\label{eq:vorticity-equation}
\end{eqnarray}

Here, $\omega_{{\rm Ds}}\equiv\omega_{{\rm ds}}(x_{\perp}^{2}/2+x_{\parallel}^{2})C$
is the magnetic drift frequency, $\omega_{{\rm ds}}\equiv k_{\theta}cT_{{\rm s}}/(q_{{\rm s}}B_{0}R)$,
$x_{\perp/\parallel}\equiv v_{\perp/\parallel}/v_{{\rm ts}}$, $v_{{\rm ts}}^{2}\equiv2T_{{\rm s}}/m_{{\rm s}}$,
$C=\cos\theta+{\rm i}k_{r}\sin\theta/k_{\theta}$ is the curvature
term, with the first and second terms representing the normal and
geodesic curvature. The first term represents the field line bending
term, and the second term is the inertial term, which enables the
coupling of $m$ and $m+1$ harmonics of TAE, i.e., linear toroidal
coupling. The last term is the curvature coupling term. It is expected
that the nonlinear CCT terms resulting from the feedback of zonal
fields and PSZS to TAE can modify the potential well of TAE, thus
causing frequency shift of TAE. The nonlinear saturation of TAE in
the cases with and without zonal fields will be discussed, respectively.

\subsection{TAE saturation without zonal fields}

In this subsection, the nonlinear saturation of TAE in the absence
of $n=0$ zonal fields is investigated, while the contribution of PSZS still exists,
corresponding to the single-$n$ simulation illustrated in sec. \ref{subsec:singlen_thermal}.
The nonlinear CCT resulting from PSZS of thermal plasma is given by
\begin{eqnarray}
{\rm CCT} & = & -\dfrac{4\pi n_{0}e^{2}}{T_{{\rm i}}}k_{\theta n}\dfrac{\omega_{{\rm di}}}{B_{0}^{2}}\dfrac{1}{\omega_{n}}\left(\dfrac{\omega_{*{\rm i}}}{\omega_{n}}-1\right)C\nonumber \\
 &  & \times\delta\phi_{n}\partial_{r}^{2}k_{\theta n}\left|\delta\phi_{n}\right|^{2}.\label{eq:CCT}
\end{eqnarray}

The two terms in the parentheses represent the contribution of thermal
ion and electron, respectively. With $\omega_{*{\rm i}}/\omega_{n}\ll1$ in mind, the contribution of the electron is much larger than that of thermal ion. Thus, the dominance of electron nonlinearity in TAE saturation in single-$n$ simulation is explained. The contribution of electron nonlinearity originates from the perturbed current due to the anti-symmetric part of PSZS, and it essentially contributes to the nonlinear potential well in the eigenmode equation of TAE, thus causing the shift in TAE frequency $\delta\omega$. Since $\delta\omega$ is positively related
to the intensity of TAE $\left|\delta\phi_{n}\right|^{2}$, as the
amplitude of TAE reaches $A_{{\rm sat}}^{{\rm T}}$, $\delta\omega$
is large enough for the mode to merge into the continuum, and TAE
becomes two independent EPMs, as observed in Figure \ref{fig:phi_s_omega_no_zf_all_nonlinearity}. Note again here that $A_{{\rm sat}}^{{\rm T}}$ is the saturation level of $e\delta\phi_n/T_{\rm e}$ due to thermal plasma nonlinearities.
Meanwhile, there exists a threshold on $\left|\delta\phi_{n}\right|^{2}$
that $\delta\omega$ is large enough to trigger the feedback loop
shown in Figure \ref{fig:schematic}, which is consistent with our
analysis given in sec. \ref{subsec:singlen_thermal}. 

The saturation level of TAE $A_{{\rm sat}}^{{\rm T}}$ can be determined
by solving the eigenmode equation of TAE with nonlinear CCT being
taken into account. However, this is beyond the scope intended here,
which would be more suitable to present in a separate publication.
Otherwise, with the weakly sheared $q$ profile adopted in this
work, the geodesic curvature in $C$ can be neglected, i.e., $C\approx\cos\theta$,
and recall the linear toroidal coupling (LTC) is proportionate to
$\epsilon\cos\theta$. Thus, the CCT can be treated as a nonlinear
decoupling term. Then, the saturation level of TAE can be roughly
estimated by simply balancing the nonlinear decoupling and LTC, which
yields the saturation level due to thermal plasma nonlinearity
\begin{eqnarray}
A_{{\rm sat,a}}^{{\rm T}} & \sim & \sqrt{2\epsilon\dfrac{1}{\beta_{\rm e}}\dfrac{k_{\parallel}^2}{k_{\theta n}^2}\dfrac{k_\parallel v_{\rm A}}{\omega_{\rm de}}}.\label{eq:thermal_sat_level}
\end{eqnarray}

Note again that $A_{{\rm sat,a}}^{{\rm T}}$ is the saturation level of $e\delta\phi_n/T_{\rm e}$ due to thermal plasma nonlinearities discussed herein. After substituting the parameters in the simulation, the saturation
level can be obtained as $A_{{\rm sat,a}}^{{\rm T}}\approx0.12$,
where the subscript ``a'' represents the analytical value. While,
the numerical value is $A_{{\rm sat}}^{{\rm T}}=0.05\sim0.1$, as
shown in Figure \ref{fig:scaling_law_no_zf}. Meanwhile, in the absence
of electron nonlinearity, the saturation level of TAE will be larger
to compensate for the smallness of the coefficient. Meanwhile, a linear
dependence of $A_{{\rm sat,a}}^{{\rm T}}$  on the square root of inverse aspect ratio
can be identified, which is consistent with our analysis in the figure. \ref{fig:sat_scaling_aspect_ratio}, indicating a larger saturation level in realistic tokamaks with
a larger inverse aspect ratio. Note again that the saturation level
given in Equation (\ref{eq:thermal_sat_level}) is only a rough estimation; detailed analysis should be based on solving the eigenmode equation
of TAE in the presence of CCT given in Equation (\ref{eq:CCT}).
By doing so, the dependence of the frequency shift on the intensity of
TAE $\left|\delta\phi_{n}\right|^{2}$ can be obtained.

\subsection{TAE saturation with zonal fields}

In the case with $n=0$ zonal fields, inserting particle responses (\ref{eq:TAE_ion_NL_full})
and (\ref{eq:TAE_electron_NL_full}) into the CCT, we have
\begin{eqnarray}
 & {\rm CCT}=\dfrac{4\pi en_{0}\omega_{\rm di}k_{\theta n}C}{B_{0}c}\dfrac{e}{T_{\rm i}}\delta\phi_{n}\partial_{r}\left(\dfrac{\omega_{n}}{\omega_{*{\rm i}}}\delta\phi_{{\rm Z}}^{{\rm T}}-\delta\phi_{{\rm Z}}\right).\label{eq:CCT_full}
\end{eqnarray}

Note that in deriving Equation (\ref{eq:CCT_full}), the contribution
of electrons still dominates over that of thermal ions; thus, the latter
is neglected. The first term is the contribution of zonal fields,
and the second term is due to PSZS. Here, $\delta\phi_{{\rm Z}}$
represents the total zonal electrostatic potential, while $\delta\phi_{{\rm Z}}^{{\rm T}}$
is that generated by thermal plasma nonlinearity, as given in Equation 
(\ref{eq:thermal_ZF}). Then, following the similar procedure, we
can obtain the saturation level of TAE $e\delta\phi_n/T_{\rm e}$
due to thermal plasma nonlinearity with zonal fields included as
\begin{eqnarray}
A_{{\rm sat,a}}^{\rm Z} & \sim & \dfrac{1}{\sqrt{1-Q}}\sqrt{2\epsilon\dfrac{1}{\beta_{\rm e}}\dfrac{k_{\parallel}^2}{k_{\theta n}^2}\dfrac{k_\parallel v_{\rm A}}{\omega_{\rm de}}},\label{eq:thermal_sat_level_with_ZF}
\end{eqnarray}
where $Q$ is defined as
\begin{eqnarray*}
Q & \equiv & \dfrac{\partial_{r}\delta\phi_{{\rm Z}}}{\partial_{r}\delta\phi_{{\rm Z}}^{{\rm T}}}\dfrac{\omega_{*{\rm i}}}{\omega_{n}}.
\end{eqnarray*}

The terms other than $1/\sqrt{1-Q}$ is the same as that without zonal
fields, i.e., Equation  (\ref{eq:thermal_sat_level}). Recall that
$\delta\phi_{{\rm Z}}\approx250\delta\phi_{{\rm Z}}^{{\rm T}}$ is
justified in Figure \ref{fig:radial_structure_ZF_ZC}, we have $Q\approx0.8$ at the position where TAE
locates. Thus, the introduction of zonal fields can result in a roughly factor of $2$
increase in the saturation level of TAE, i.e., from $e\delta\phi_n/T_{\rm e}\approx0.12$
in the case without zonal fields to $e\delta\phi_n/T_{\rm e}\approx0.27$ in the case with zonal fields, which quantitatively agrees with the simulation
result given in Figure \ref{fig:with_without_ZF_compare_phimax}.
Thus, the rough estimation given above can, at least, explain the
simulation observations. However, as we mentioned earlier, a more precise analysis should be carried out in ballooning space to find
out the nonlinear modifications of zonal fields and PSZS to the potential well
of TAE. 

In fact, zonal fields and PSZS must be generated simultaneously; thus, the case studied here with both zonal fields and PSZS is a self-consistent system. On the
contrary, the single-$n$ simulation  with zonal fields filtered out and PSZS of thermal species retained is an imbalanced system, leading to a more stabilized zonal
state and an underestimated saturation level of TAE. The introduction
of zonal fields balances the effects of
PSZS, resulting in a more unstable zonal state and a higher saturation
level of TAE. This implies that for PIC codes, $n=0$ modes should
be retained when thermal plasma evolves nonlinearly, in order
to avoid the imbalance and obtain a more self-consistent result. 

The ITPA case studied here is a case with very weak magnetic shear
and a large aspect ratio with $a/R=0.1$. Both Equations (\ref{eq:thermal_sat_level}) and (\ref{eq:thermal_sat_level_with_ZF}) predict a positive dependence on the square root of inverse aspect ratio. 
Figure \ref{fig:sat_scaling_aspect_ratio} shows the scaling law for $a/R=0.1$ and $a/R=0.2$, which indicates that the saturation level of TAE is $\sqrt{2}$ times larger when the inverse aspect ratio is doubled for both cases with and without zonal fields, which is consistent with our theoretical prediction. This is different from the $(a/R)^{5/2}$ dependence in Ref. \cite{FZoncaPRL1995} and $(a/R)^{3/2}$ in Ref. \cite{LChenPPCF1998}.
Nevertheless, the results given in these scalings represent the fact that in realistic cases with a larger inverse aspect ratio, the saturation level of TAE will be
larger. Note here that Equation (\ref{eq:thermal_sat_level_with_ZF}) only holds for $Q<1$, and for realistic
cases with larger $\omega_{*{\rm i}}$, the saturation level should
be obtained by solving the eigenmode equation of TAE, which might be presented in a future publication.

\begin{center}
\begin{figure}
\begin{centering}
\includegraphics[scale=0.4]{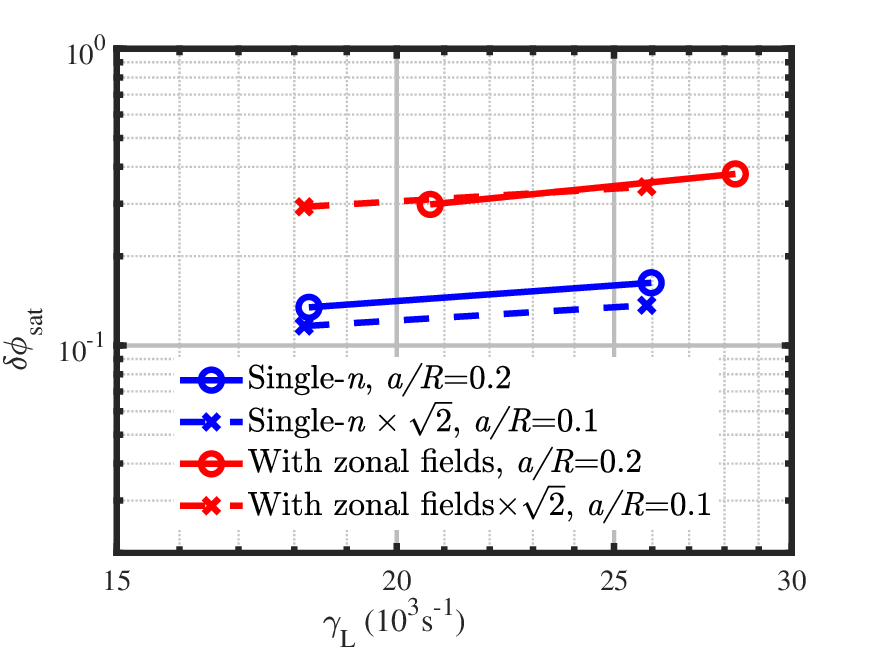}
\par\end{centering}
\caption{The dependence of saturation level of TAE on the linear drive for $a/R=0.1$ (dashed lines) and $a/R=0.2$ (solid lines) in both cases with zonal fields (red lines) and without zonal fields (blue circles). Here, the saturation level in single-$n$ simulation is multiplied by $\sqrt{2}$.}\label{fig:sat_scaling_aspect_ratio}
\end{figure}
\par\end{center}

\section{Summary and Discussion}\label{sec:discussion}

In this work, we have carried out a series of simulations of toroidal Alfv{\'e}n eigenmode (TAE) based on the standard International Tokamak Physics Activity (ITPA) set of parameters to investigate the effects of thermal
plasma nonlinearities on nonlinear saturation of TAE in the drift-kinetic
limit for both cases with and without zonal fields. Meanwhile, we
have developed a theoretical model to explain the simulation findings,
and good agreements are identified. For the cases with both $n=0$ and $6$
modes, there is $n=0$ zonal fields and phase-space zonal structures
(PSZS) for each particle species generated by beating of $n=6$ TAE
and itself. The universality of zonal fields and PSZS in the presence of finite level TAE results from the thresholdless nature of the beat-driven process. However, in numerical simulations,  $n=0$ zonal fields can be systematically filtered out, i.e., single-$n$ simulation given in this paper. Meanwhile, PSZS can not be filtered out in particle-in-cell (PIC) codes when the corresponding particle species evolves nonlinearly. 

In the simulations without $n=0$ zonal fields, zonal fields are systematically filtered out, while there always exists PSZS in the system. It is found that the inclusion of thermal plasma nonlinearity
imposes a nearly constant threshold on the saturation level of TAE $e\delta\phi_{n}/T_{{\rm e}}=0.05\sim0.1$ for linear drive $\gamma_{\rm L}/\omega_{n}>0.47\%$, i.e., ``stiffness'' in saturation.
The saturation of TAE due to thermal plasma nonlinearities can be reached even if EPs evolve linearly, i.e., constant EP drive.
The mechanism for saturation is found to be the frequency downshift, which can result
in strong coupling with the continuum.
Upon saturation, the neighboring $m=10$ and $m=11$ harmonics of
TAEs are decoupled, leading to a mode transition to energetic particles
modes (EPMs). The gyrokinetic theory developed here indicates that the frequency downshift is resulted from the PSZS of thermal species. Then, a rough estimation on the saturation level can be obtained by balancing the linear toroidal
coupling and nonlinear decoupling induced by PSZS of thermal plasmas,
which agrees quantitatively with simulation results. 

In the simulations with $n=0$ zonal fields retained, we found that zonal fields can significantly balance the effects of PSZS on mode saturation, leading to 
a roughly factor of $2$ enhancement to the saturation level of TAE. The saturation is dominated by thermal plasma nonlinearity for reasonable linear growth rate $\gamma_{\rm L}/\omega_{n}>1\%$. In both cases with and without zonal fields, the gyrokinetic theory predicts a positive dependence of the saturation level of TAEs on the square root of inverse aspect ratio, which is consistent  with simulation results. 

Above all, the results illustrated above imply that $n=0$ zonal fields should be retained to obtain a self-consistent
 saturation level of TAE in PIC simulations, or the saturation level will be underestimated if only PSZS is retained in the system. Meanwhile, the positive dependence of
the saturation level due to thermal plasma nonlinearities on the square root of inverse aspect ratio possibly indicates a larger saturation level for realistic
 tokamaks with a large inverse aspect ratio. However, both numerical
and theoretical analysis given in this work is based on a specific
large aspect ratio ITPA configuration; thus, the extension to more
realistic cases remains to be investigated.

\section*{Acknowledgement}

The authors acknowledge Professor Liu Chen (Zhejiang University, PRC) for inspiration on physical understanding and theoretical analysis.
This work has been carried out within the framework of the EUROfusion Consortium, funded by the European Union via the Euratom Research and Training Programme (Grant Agreement No 101052200—EUROfusion), and the International Partnership Program of Chinese Academy of Sciences (Grant No. 145GJHZ2025017BS). Views and opinions expressed are however those of the author(s) only and do not necessarily reflect those of the European Union or the European Commission. Neither the European Union nor the European Commission can be held responsible for them. 

\section*{Appendix A. Convergence study}

The convergence is mainly investigated in terms of the mass ratio, and the convergence in terms of time steps and markers is trivial, so this will not be presented here. For the single-$n$ simulations, the convergence can be reached with mass ratio being $m_{\rm i}/m_{\rm e}=200$ and time step being $\omega_{\rm ci}\Delta t=20$,  marker numbers being $10^7,4\times10^7,4\times10^7$ for thermal ions, electrons, EPs, which is consistent with Ref. \cite{ABiancalaniPPCF2017}.

\begin{center}
\begin{figure}
\begin{centering}
\includegraphics[scale=0.4]{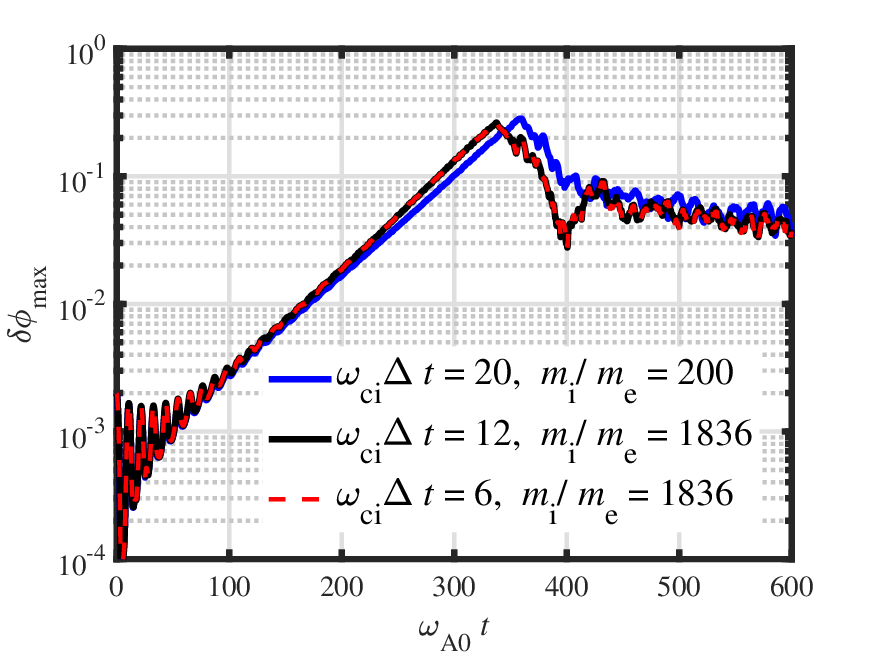}
\par\end{centering}
\caption{The temporal evolution of $e\delta\phi_n/T_{\rm e}$ for different combinations of time step and mass ratio. Here, the EP concentration $n_{\rm EP}/n_{\rm e}=0.002$.}\label{fig:convergence_mass_ratio}
\end{figure}
\par\end{center}

\begin{center}
\begin{figure}
\begin{centering}
\includegraphics[scale=0.4]{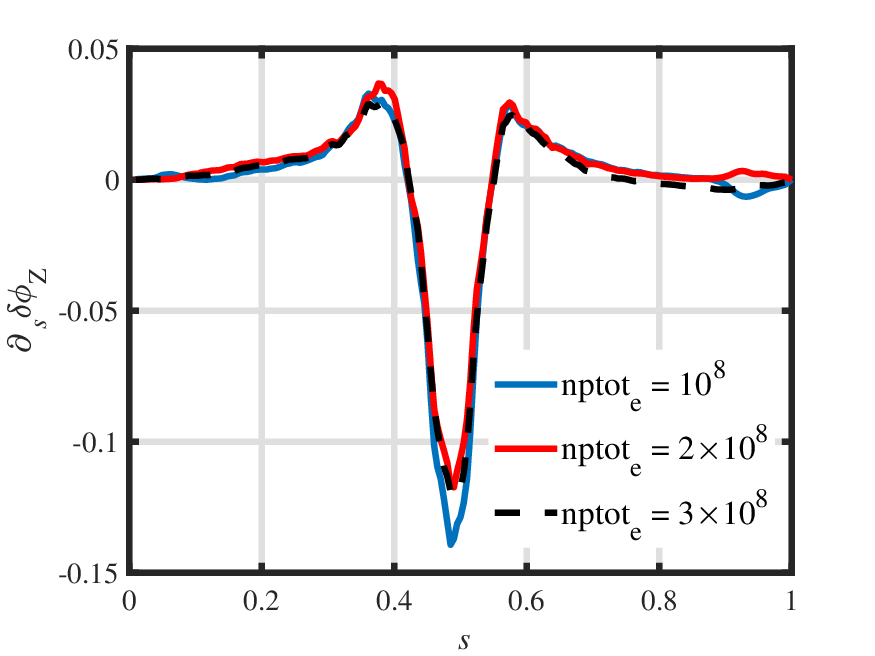}
\par\end{centering}
\caption{The radial mode structure of $\partial_s\delta\phi_{\rm Z}$ for different electron marker numbers. Here, the mass ratio is $m_{\rm i}/m_{\rm e}=1836$, time step is $\omega_{\rm ci}\Delta t=12$, and the EP concentration is $n_{\rm EP}/n_{\rm e}=0.002$.}\label{fig:convergence_nptot}
\end{figure}
\par\end{center}

For the case with zonal fields, the convergence with respect to mass ratio and time step is shown in Figure \ref{fig:convergence_mass_ratio}, indicating the convergence can be reached with $m_{\rm i}/m_{\rm e}=1836$ and $\omega_{\rm ci}\Delta t=12$. Meanwhile, as we have both $n=0$ and $n=6$ modes, the number of markers, especially electron markers, should increase to distinguish two modes. Thus, a scan in electron marker numbers is performed with radial mode structure of $\partial_s\delta\phi_{\rm Z}$, which is shown in Figure \ref{fig:convergence_nptot}. The convergence can be found with the electron marker number being $2\times10^8$. Above all, the default settings for the cases with zonal fields are $2\times10^7,2\times10^8,4\times10^7$ for thermal ions, electrons, EPs, $m_{\rm i}/m_{\rm e}=1836$, $\omega_{\rm ci}\Delta t=12$ for $n_{\rm EP}/n_{\rm e}=0.002$. The marker numbers will increase as the linear drive decreases in order to get a better signal-to-noise ratio. Finally, it turns out that the computational cost needed for the case with zonal fields will be an order of magnitude larger than that without zonal fields, resulting in the difficulty with exploring into the weak linear drive regime.

\bibliography{reference}

\end{document}